\documentclass[sigconf, nonacm, pdfa]{acmart}

\newcommand\vldbdoi{10.14778/3611540.3611548}
\newcommand\vldbpages{3570 - 3583}
\newcommand\vldbvolume{16}
\newcommand\vldbissue{12}
\newcommand\vldbyear{2023}
\newcommand\vldbauthors{\authors}
\newcommand\vldbtitle{\shorttitle} 
\newcommand\vldbavailabilityurl{}
\newcommand\vldbpagestyle{empty}

\usepackage[a-2b]{pdfx}

\usepackage{balance}
\usepackage{mathtools}

\usepackage{subfigure}
\usepackage{multirow}
\usepackage{bm}
\usepackage{bbm}
\usepackage{dsfont}
\usepackage{color,soul}
\usepackage{url}
\usepackage{listings}
\usepackage{xcolor}
\usepackage{algorithmic}

\usepackage{natbib}

\usepackage[linesnumbered,vlined,ruled,noend]{algorithm2e}

\usepackage{enumitem}
\usepackage{booktabs}
\usepackage{caption}
\usepackage{bm}

\begin{document}

\title{Towards General and Efficient Online Tuning for Spark}



\author{Yang Li}
\affiliation{
  \institution{Tencent Inc.}
}
\email{thomasyngli@tencent.com}

\author{Huaijun Jiang}
\affiliation{%
  \institution{Peking University \& Tencent Inc.}
}
\email{jianghuaijun@pku.edu.cn}

\author{Yu Shen}
\affiliation{%
  \institution{Peking University}
}
\email{shenyu@pku.edu.cn}

\author{Yide Fang}
\affiliation{%
  \institution{Tencent Inc.}
}
\email{yidefang@tencent.com}

\author{Xiaofeng Yang}
\affiliation{%
  \institution{Tencent Inc.}
}
\email{felixxfyang@tencent.com}

\author{Danqing Huang}
\affiliation{%
  \institution{Tencent Inc.}
}
\email{daisyqhuang@tencent.com}

\author{Xinyi Zhang}
\affiliation{%
  \institution{Peking University}
}
\email{zhang\_xinyi@pku.edu.cn}

\author{Wentao Zhang}
\affiliation{%
  \institution{Mila - Qu\'ebec AI Institute}
}
\email{wentao.zhang@mila.quebec}

\author{Ce Zhang}
\affiliation{%
  \institution{ETH Z\"urich}
}
\email{ce.zhang@inf.ethz.ch}

\author{Peng Chen}
\affiliation{
  \institution{Tencent Inc.}
}
\email{pengchen@tencent.com}

\author{Bin Cui}
\affiliation{%
  \institution{Peking University}
}
\email{bin.cui@pku.edu.cn}








\renewcommand{\shortauthors}{Li et al.}

\renewcommand{\authors}{Yang Li, Huaijun Jiang, Yu Shen, Yide Fang, Xiaofeng Yang, Danqing Huang, Xinyi Zhang, Wentao Zhang, Ce Zhang, Peng Chen, Bin Cui}

\begin{abstract}
The distributed data analytic system -- Spark is a common choice for processing massive volumes of heterogeneous data, while it is challenging to tune its parameters to achieve high performance.
Recent studies try to employ auto-tuning techniques to solve this problem but suffer from three issues: limited functionality, high overhead, and inefficient search.

In this paper, we present a general and efficient Spark tuning framework that can deal with the three issues simultaneously.
First, we introduce a generalized tuning formulation, which can support multiple tuning goals and constraints conveniently, and a Bayesian optimization (BO) based solution to solve this generalized optimization problem.
Second, to avoid high overhead from additional offline evaluations in existing methods, we propose to tune parameters along with the actual periodic executions of each job (i.e., online evaluations).
To ensure safety during online job executions, we design a safe configuration acquisition method that models the safe region.
Finally, three innovative techniques are leveraged to further accelerate the search process: adaptive sub-space generation, approximate gradient descent, and meta-learning method.

We have implemented this framework as an independent cloud service, and applied it to the data platform in Tencent. 
The empirical results on both public benchmarks and large-scale production tasks demonstrate its superiority in terms of practicality, generality, and efficiency. 
Notably, this service saves an average of 57.00\% memory cost and 34.93\% CPU cost on 25K in-production tasks within 20 iterations, respectively.

\end{abstract}

\maketitle

\pagestyle{\vldbpagestyle}
\begingroup\small\noindent\raggedright\textbf{PVLDB Reference Format:}\\
\vldbauthors. \vldbtitle. PVLDB, \vldbvolume(\vldbissue): \vldbpages, \vldbyear.\\
\href{https://doi.org/\vldbdoi}{doi:\vldbdoi}
\endgroup
\begingroup
\renewcommand\thefootnote{}\footnote{\noindent
This work is licensed under the Creative Commons BY-NC-ND 4.0 International License. Visit \url{https://creativecommons.org/licenses/by-nc-nd/4.0/} to view a copy of this license. For any use beyond those covered by this license, obtain permission by emailing \href{mailto:info@vldb.org}{info@vldb.org}. Copyright is held by the owner/author(s). Publication rights licensed to the VLDB Endowment. \\
\raggedright Proceedings of the VLDB Endowment, Vol. \vldbvolume, No. \vldbissue\ %
ISSN 2150-8097. \\
\href{https://doi.org/\vldbdoi}{doi:\vldbdoi} \\
}\addtocounter{footnote}{-1}\endgroup

\ifdefempty{\vldbavailabilityurl}{}{
\vspace{.3cm}
\begingroup\small\noindent\raggedright\textbf{PVLDB Artifact Availability:}\\
The source code, data, and/or other artifacts have been made available at \url{\vldbavailabilityurl}.
\endgroup
}

\section{Introduction}
\label{sec:introduction}
The rapid growth of the World Wide Web, E-commerce, Social media and other applications is producing massive amounts of ever-increasing raw data every day~\cite{herodotou2020survey}. 
Companies often utilize this ``big data'' to innovate pioneering products and solutions, e.g., improving customer service, marketing, sales, team management, and many other routine operations.
To this end, many distributed analytics platforms (e.g., Hadoop Mapreduce~\cite{dean2008mapreduce,dittrich2012efficient}, Spark~\cite{zaharia2012resilient,armbrust2015spark,zaharia2016apache}, Storm~\cite{toshniwal2014storm}, Flink~\cite{carbone2015apache}, Heron~\cite{kulkarni2015twitter}, Samza~\cite{noghabi2017samza}) have emerged to deal with this big data trend.
Spark is one of the representative systems that enable the manipulation and analysis of large datasets with in-memory cluster computing~\cite{zaharia2016apache}.
As of today, thousands of organizations~\cite{sparkwebsite}, such as Google, Microsoft, Amazon, Meta, Oracle, Snowflake, Databricks, Tencent, and Alibaba are using Spark in production across a vast range of fields including data processing~\cite{doulkeridis2014survey,lee2012parallel}, machine learning~\cite{meng2016mllib}, graph computing~\cite{xin2013graphx}, stream computing~\cite{sparksteamingwebsite} and database management~\cite{armbrust2015spark,xin2013shark}.

The performance of Spark jobs highly depends on the choice of Spark configuration parameters.
Misconfiguration can lead to unsatisfying performance such as long runtime, resource contention, and resource under-utilization~\cite{alipourfard2017cherrypick}. 
For instance, an improper configuration may lead to over 100 times the execution time compared with an elaborately designed one~\cite{yu2018datasize}.
Therefore, it is crucial to choose proper configurations to achieve high performance (e.g., in terms of execution time or cost).
The benefit of tuning is even more significant for periodic (hourly, daily, weekly, etc) jobs, where they occupy a large proportion of Spark jobs in production~\cite{alipourfard2017cherrypick}. In this paper, we focus on tuning periodic Spark jobs.

To achieve a near-optimal performance of periodic tasks, users are required to determine a large number of performance-critical configuration parameters.
However, it is very difficult to manually tune the configuration parameters of a Spark task due to (1) the high-dimensionality of the parameter space and (2) complex and non-linear interactions among parameters.
In addition, manual tuning is usually time-consuming and labor-intensive, and thus it fails to scale to a huge number of tasks in a data platform.
Recent studies propose to utilize auto-tuning techniques to optimize the Spark parameters. 
During our attempts to apply these approaches in real tuning tasks, we realize three aspects of limitations:

{\bf (C.1 Limited Functionality)} Lots of methods (e.g., DAC~\cite{yu2018datasize}, LOCAT~\cite{locat}) are designed to minimize the execution time, i.e., finding the fastest configuration.
However, the goal in many scenarios involves the execution cost~\cite{alipourfard2017cherrypick,shen2023rover} (i.e., the cheapest configuration), or more generalized objectives such as a weighted combination between runtime and cost.
In addition, the tuning process should satisfy some application requirements. 
For instance, the execution time of a configuration should be lower than a given threshold~\cite{alipourfard2017cherrypick}.
These scenarios call for a generalized formulation and framework for tuning, which is not supported by existing methods.

{\bf (C.2 High Overhead)}
Many tuning frameworks~\cite{herodotou2011starfish,lama2012aroma,bei2015rfhoc,yu2018datasize} belong to the offline tuning paradigm.
Concretely, they collect training samples by running jobs with different configurations on a non-production cluster, and then train a performance model to suggest new configurations for Spark job running on the production cluster.
Training such an accurate performance model involves lots of offline job executions (e.g., 1000-10000)~\cite{locat}.
This process is very time-consuming and expensive, and often incurs data security issues when accessing business data during multiple job executions in the non-production cluster.
In addition, since workloads (e.g., data size) may change as time proceeds~\cite{yu2018datasize}, offline tuning methods cannot capture and adapt to these dynamics easily.

{\bf (C.3 Inefficient Search)}
The configuration search in many methods~\cite{yu2018datasize,alipourfard2017cherrypick} suffers from the low-efficiency issue that arises from two aspects: 
(1) huge search space and (2) intrinsic dilemma of black-box optimization.
As dimensionality increases, the number of samples required to learn the performance model grows exponentially (i.e., the curse of dimensionality).
Therefore, these methods fail to provide quick
convergence in high-dimensional space.
Furthermore, this tuning task, in essence, the black-box optimization problem, faces the cold-start issue and challenge when dealing with the exploration and exploitation trade-off.
These factors greatly affect the convergence speed.
In addition, as studied in optimization community~\cite{feurer2018scalable,golovin2017google}, meta knowledge across tasks could help to deal with cold-start and slow convergence issues during search~\cite{li2022volcanoml,bai2023transfer,li2022transfer,li2022transbo}.
However, nearly none of the existing methods utilize the tuning history from previous tasks to accelerate the current tuning process.


\begin{table}[tb]
    \caption{Summary of several tuning methods.
    \textit{General obj.} notes whether the method supports generalized objectives or not, e.g., runtime or cost. 
    \textit{Constr.} indicates if it supports performance constraints.
    \textit{NOER} refers to the ``No Offline Evaluation Required'', which usually causes high overhead. 
    \textit{Safety} represents the ability to achieve safe config. acquisition during search.
    \textit{Adaptive space} notes if it can adjust the search space adaptively to deal with the high-dimensional issue.
    \textit{Meta-learn} indicates whether it utilizes meta-learning techniques to achieve quick convergence.
    $\triangle$ means the method partially supports it with some limitations.}
    \label{tbl:method_cmp}
    \scriptsize
    \centering
    \begin{tabular}{l|ccccccc}
        \hline
        Method & General obj. & Constr. & NOER & Safety & Adaptive space & Meta-learn \\
        \hline
        RFHOC & $\times$ & $\times$ & $\times$ & $\times$ & $\times$ & $\times$ \\
        DAC & $\times$ & $\times$ & $\times$ & $\times$ & $\times$ & $\times$ \\
        CherryPick & $\times$ & $\triangle$ & \checkmark & $\times$ & $\times$ & $\times$ \\
        Tuneful & $\times$ & $\times$ & \checkmark & $\times$ & $\triangle$ & \checkmark \\
        LOCAT & $\times$ & $\times$ & \checkmark & $\times$ & $\triangle$ & $\times$ \\
        Ours & \checkmark & \checkmark & \checkmark & \checkmark & \checkmark & \checkmark \\
        \hline
    \end{tabular}
\end{table}

To address these issues simultaneously (See Table~\ref{tbl:method_cmp}), we present a general online tuning framework for Spark, which can efficiently unearth the optimal/near-optimal configurations that minimize the objectives under different scenarios (e.g., runtime or cost) and satisfy application requirements.
Concretely, to deal with limited functionality (\textbf{C.1}), we first introduce a {\bf generalized formulation} about the tuning problem, which supports various goals and multiple performance constraints conveniently. 
To solve this generalized optimization problem, we develop a noise-robust Bayesian optimization (BO)~\cite{shahriari2015taking} based solution.
Second, considering the high overhead issue (\textbf{C.2}), different from offline approaches that require a large number of job executions in advance, we propose an {\bf online tuning paradigm}, where we tune parameters along with the in-production periodic executions of each Spark job, thus incurring no additional execution cost as in offline methods.
This online setting requires that the BO-based framework should converge to good configurations quickly (efficiency) and explore as few bad configurations as possible to get there (safety).
To this end, we design a configuration acquisition method, which models and utilizes the safe region from the Gaussian process~\cite{rasmussen2003gaussian} to achieve safe exploration, and then leverages the expected constrained improvement to trade-off exploration and exploitation when suggesting new configurations. 
To accommodate the dynamic workload in the online setting, we encode the workload characteristic into the Gaussian process using mixed kernels in BO.

As for the inefficient search issue (\textbf{C.3}),
we develop three innovative techniques within BO to accelerate tuning.
First, Bayesian optimization is known to be difficult to scale to high dimensions.
To overcome this high-dimensionality issue, we propose an {\bf adaptive sub-space generation} method to obtain a much smaller space for configuration search, instead of the original huge space. 
The size of the sub-space will be automatically adjusted based on the intermediate tuning results to balance convergence speed and quality.
Second, since gradient methods~\cite{ruder2016overview} have shown superior convergence speed compared with black-box methods in numerical optimization, we develop a novel {\bf approximate gradient descent} method within BO to select the next configuration, which can estimate and leverage the derivative information of the objective function to speed up the search process.
Third, motivated by the observation that the regions of optimal/near-optimal configurations are similar between two similar tuning tasks, we design a {\bf meta-learning module} that learns the similarity between tasks based on tuning history from previous tasks.
Further, this module can transfer useful knowledge, e.g., the optimal configuration or the distribution of good configurations, to the current tuning task.
Therefore, it takes fewer configuration trials to find the near-optimal configuration.

{\bf (Contributions)}
The main contributions of this paper can be summarized as follows:
{\bf (1)} We propose a general and efficient online tuning framework for Spark. To our knowledge, it is the first framework that can deal with the three issues simultaneously. 
{\bf (2)} To solve the generalized tuning problem while pursuing safety in the online setting, we design a Bayesian optimization (BO) based solution along with a safe configuration acquisition approach.
{\bf (3)} To further accelerate configuration search, we develop three innovative techniques within BO: adaptive sub-space generation, approximate gradient descent, and meta-learning-based knowledge transfer.
{\bf (4)} We have implemented this tuning framework as a cloud service, and applied it to the Tencent data platform. 
The empirical results on public benchmarks and large-scale production tasks demonstrate its generality and efficiency when tuning different objectives.
Notably, compared with manual settings, our service saves 57.00\% and 34.93\% of memory and CPU cost on average over 25K in-production Spark tasks within 20 iterations, respectively.

\begin{figure*}[t!]
	\centering
	\scalebox{0.9}[.9] {
	\includegraphics[width=1\linewidth]{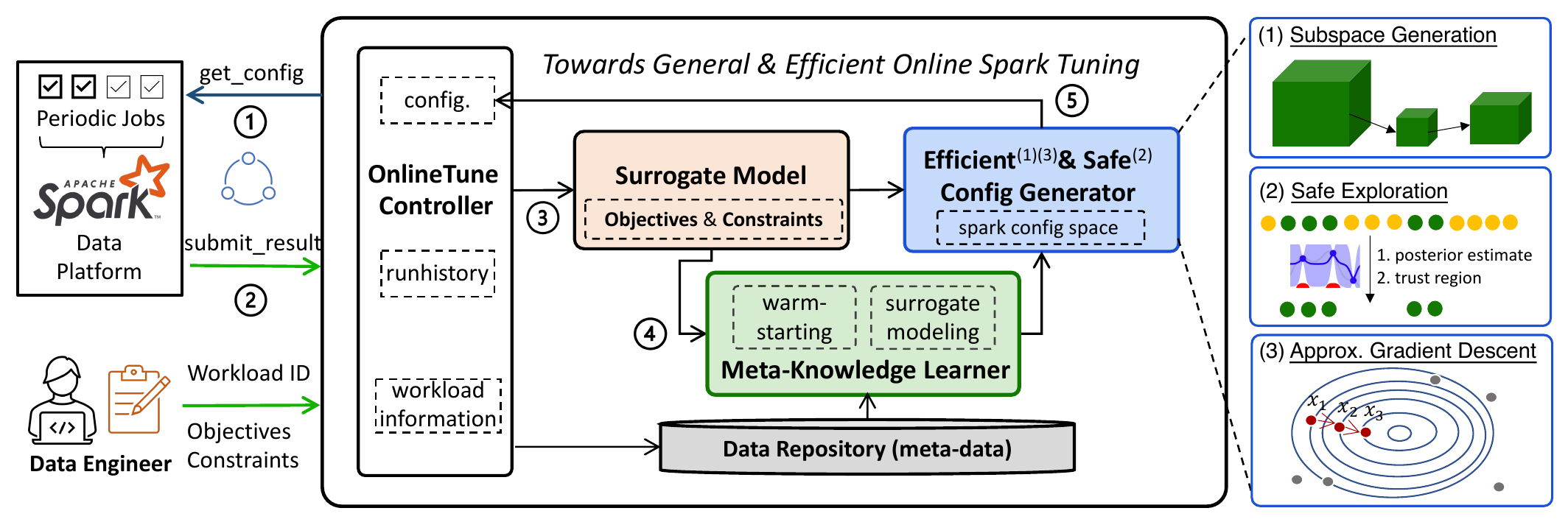}
         }
	\caption{Online Spark tuning architecture and interaction with Spark.}
    \label{fig:framework}
\end{figure*}

\section{Background}
\subsection{Spark Framework}

Spark is an open-source distributed framework, which simplifies the development of applications that execute in parallel on computing clusters. 
Spark stores the data in memory as Resilient Distributed Datasets (RDDs), and this design choice significantly reduces I/O costs and speeds up iterative job execution time.
RDDs are immutable collections distributed over a cluster of machines and maintained in a fault-tolerant way.
Spark provides two types of programming abstractions: transformations and actions, to manipulate RDDs.
When a Spark application (job) is submitted, the Spark framework generates a directed acyclic graph (DAG) based on the RDD dependency behind the Spark program.
Subsequently, the DAG is split into a collection of stages with each containing a set of parallel tasks. 
Each task, one per RDD partition, computes partial results of a Spark job. 
Next, the DAG scheduler schedules the stages to execute, and the cluster manager assigns the tasks to run on executors in parallel. 

The resource that can be used by an executor is specified by configuration parameters.
For example, \texttt{spark.executor.memory} and \texttt{spark.executor.cores} determine the memory size and the number of CPU cores allocated for an executor, respectively. 
A Spark application is controlled by up to 160 configuration parameters, which determine many aspects including dynamic allocation, scheduling, shuffle behavior, data serialization, memory management, execution behavior, networking, etc.
For details about more Spark parameters, please refer to the official documentation~\cite{spark_doc}.
Some parameters such as \texttt{spark.application.name} do not affect performance.
Following previous work~\cite{fekry2020tuneful}, in this paper, we focus on tuning 30 parameters that significantly affect job performance. 

\subsection{Terminology}

\noindent\textbf{Configuration Space.} 
Our goal is to find the optimal or near-optimal configuration from the configuration space, which contains $N$ Spark parameters: $x_{p1}, x_{p2}, ..., x_{pN}$.
Each $p_i$ corresponds to a specific parameter in Spark, e.g., \texttt{spark.executor.instances}, \texttt{spark.executor.memory}.
The range of the $i^{th}$ parameter $p_i$ is denoted by $\Lambda^{i}$ and the whole configuration space can be represented by $\bm{\Lambda}_{cs}=\Lambda^1 \times \Lambda^2 \times ... \Lambda^N$.
Note that, we use $\bm{x}$ to denote a configuration instance (a vector) in $\bm{\Lambda}_{cs}$, where the $i^{th}$ element $\bm{x}^i$ corresponds to the value of $i^{th}$ parameter $p_i$ and $x^i \in \Lambda^i$.

\noindent\textbf{Configuration Subspace.}
A configuration subspace $\bm \Lambda_{sub}$ is the set of all possible combinations of values of each parameter in the space, which only includes a part of Spark parameters. 
Therefore, the size of the configuration subspace usually is much smaller than the original configuration space $\bm \Lambda_{cs}$.

\noindent\textbf{Tuning Procedure.}
Given a configuration space and tuning budget, a tuner applies a certain strategy to navigate the configuration space.
The goal of the tuning task is to find the optimal configuration that minimizes the underlying objective.
Common tuning strategies include random search~\cite{bergstra2012random}, grid search, genetic algorithm~\cite{bei2015rfhoc,yu2018datasize,olson2019tpot}, Bayesian optimization~\cite{alipourfard2017cherrypick,snoek2012practical,bergstra2011algorithms,hutter2011sequential}, etc.
In each iteration, the tuner suggests a configuration, and then a job execution is launched to evaluate this configuration.
The budget is some kind of \textit{cost} that is allowed to tune a Spark job; here, the tuning budget is set to the number of iterations.
Most existing approaches~\cite{bei2015rfhoc,yu2018datasize,herodotou2020survey} are designed to optimize execution time only or resort to the offline tuning paradigm. 
In the following, we investigate the generalized and efficient solutions for online Spark tuning.

\section{System Design}
In this section, we introduce the architecture overview, generalized problem formulation, and base solution framework in turn.

\subsection{Overview}
\label{sec:overview}
Figure \ref{fig:framework} presents the workflow of our framework.
The framework contains five components:
(1) the OnlineTune controller is responsible for orchestrating the entire configuration tuning process, along with interactions with the data platform and end users.
(2) The multi-purpose surrogate models are to learn complex relationships between configurations and objective metrics or performance constraints.
(3) The efficient and safe configuration generator is to suggest a promising configuration to evaluate for a tuning task.
(4) The meta-knowledge learner is capable of leveraging tuning history from previous tasks to further accelerate the search for configurations.
(5) The data repository is to store tuning-related data, including runhistory, workload metrics, etc.

To define a tuning task for a Spark job, end users first need to specify the objective and constraints they want to optimize or restrict, and set the overall optimization budget.
Then, the iterative tuning workflow is activated.
Before the execution of a periodic job starts, the data platform requests the {\bf OnlineTune controller} to pass a new configuration for this run (Step 1 in Figure~\ref{fig:framework}). 
After the job execution (i.e, the evaluation of the corresponding configuration) is finished, the data platform submits the execution results to the controller (Step 2 in Figure~\ref{fig:framework}). 
This tuning process can be stopped or resumed when the stopping or restarting  condition is reached.
When the tuning is stopped, the controller returns the best configuration found during tuning.
{\em Note that, this online tuning procedure does not require any additional job executions (i.e., configuration evaluations) as in offline methods.}

Once the OnlineTuner controller receives a configuration request from the data platform, it starts to train the {\bf surrogate models} (in Section~\ref{sec:bo_surrogate}) on the runhistory related to the current task (Step 3 in Figure~\ref{fig:framework}); the surrogates consider parameter configurations, objectives, constraints and workload characteristics.
Then, based on the surrogates, the {\bf efficient and safe configuration generator} (in Section~\ref{sec:config_acquisition}) elaborately selects the next configuration over a safe and concise sub-space, together with an efficient approximate gradient descent method (Step 5 in Figure~\ref{fig:framework}).
Finally, the configuration is sent out to the OnlineTune controller as the response to the configuration request from the data platform.
After the job equipped with the configuration is finished, the controller receives its execution results (i.e., evaluation metrics and workload information) and stores them in the \textbf{data repository}. 
During the above workflow, the {\bf meta-knowledge learner} (in Section~\ref{sec:meta-learning}) aids the configuration generator to efficiently identify and suggest promising configurations (Step 4 in Figure~\ref{fig:framework}) by leveraging the auxiliary knowledge from the tuning history of previous tasks, which is stored in the data repository.
This meta-learner could greatly decrease the number of runs to find the near-optimal configurations.

\subsection{Generalized Problem Formulation}
\label{sec:problem}
Given a tuning task, our goal is to find the optimal or near-optimal Spark configuration that minimizes the objective and satisfies performance/safety requirements. 
To support general Spark tuning cases, we provide a generalized tuning formulation that supports both different objectives and inequality constraints from real applications.
Formally, we use $T(\bm{x})$ to denote the runtime function for the tuning task. 
The runtime depends on the configuration vector $\bm{x}$, which includes resource configurations and other execution-related configurations. 
Let $P(\bm{x})$ be the price per unit of time for all computing resources used in configuration $\bm{x}$. 
Due to the unknown hardware types, VM types, changeable pricing policy, etc, 
given a $\bm{x}$, the price per unit time (i.e, $P(\bm{x})$) is not easy to obtain.
Note that, the price $P(\bm{x})$ is positively correlated with the amount of resource used in $\bm{x}$ given a fixed computing environment.
For simplicity, we use the resource function $R(\bm{x})$ to replace $P(\bm{x})$, which indicates the amount of resource used by the job execution with configuration $\bm{x}$.
Further, we formulate this problem as follows:
\begin{equation}
\small
    \label{eq:def:gentuning}
    \begin{aligned}
        & \underset{{\bm x \in \bm \Lambda_{cs}}}{\text{minimize}}
        & & f({\bm x}) = T(\bm x)^\beta \times R(\bm x)^{1-\beta}
        \\
        & \text{s.t.}
        & & T(\bm{x}) \le T_{max}, R(\bm{x}) \le R_{max}, etc\\
    \end{aligned}
\end{equation}
where $f(\bm{x})$ is the objective result of Spark configuration $\bm{x}$, $T_{max}$ and $R_{max}$ is the maximum tolerated runtime and resource used, and $\beta$ is a constant used to control the tuning goal with $\beta \in [0, 1]$. 
Resource function $R(x)$ indicates the resource used during the job execution using configuration $\bm{x}$ in terms of CPU cores and memory, and thus can be obtained directly based on the values of resource-related parameters (e.g., \texttt{spark.executor.instances}, \texttt{spark.executor.cores}, \texttt{spark.executor.memory}).

\noindent
{\bf Generalized Objective with Different $\beta$. }
Users can customize their tuning objectives and reflect their application-specific tendency by setting different values of $\beta$.
When $\beta$ is set to $1$, the target is to find the configuration with the lowest runtime; when $\beta$ equals $0$, the objective becomes minimizing the unit of resource used.
Particularly, when $\beta=0.5$, the objective is equivalent to optimizing the execution cost by ignoring the square root.
The execution cost is the product of runtime $T(\bm{x})$ and the price per unit time $P(\bm{x})$.
For simplicity, we utilize the amount of resource $R(\bm{x})$ to approximate $P(\bm{x})$ because $P(\bm{x})$ is usually proportional to $R(\bm{x})$ and is hard to determine.
If $P(\bm{x})$ can be accessed easily, users can use $P(\bm{x})$ directly in the definition of the objective function, instead of $R(\bm{x})$.
Besides, the other choices of $\beta$ could express users' objective tendencies.
For instance, $\beta=0.7$ indicates that the user pays more attention to the decrease in runtime while minimizing the execution cost.

\noindent
{\bf Application Requirements. }
The generalized formulation also supports different constraints, which is essential to ensure the quality and safety of online Spark jobs. 
As shown in Eq.~\ref{eq:def:gentuning}, users can specify the requirements about runtime, the amount of resource, etc. 
In addition, this framework allows users to customize different requirements for their tuning tasks. 
For example, the usage of CPU cores should be lower than a certain threshold.

Knowing $T(\bm{x})$ under all configurations would make it straightforward to solve Eq.~\ref{eq:def:gentuning}, but it is infeasible because all candidates need to be evaluated.
Instead, we use the Bayesian optimization (BO) based method to directly search for an approximate solution of Eq.~\ref{eq:def:gentuning} with significantly fewer online job executions.

\subsection{Bayesian Optimization-based Framework}
\label{sec:bo_surrogate}

\begin{algorithm}[t]
  \small
  \caption{Pseudo code for Bayesian optimization.}
  \label{algo:bo}
  \begin{algorithmic}[1]
  \REQUIRE the search budget $\mathcal{B}$, the configuration search space $\bm \Lambda_{cs}$.
  \ENSURE the best configuration found.
  \STATE initialize observations $D$ with several configuration evaluations.
  \WHILE{budget $\mathcal{B}$ does not exhaust} 
  \STATE train a surrogate $M$ on the current observations $D$.
  \STATE choose the next configuration by $\bm{x}_{i}=\arg\max_{\bm{x} \in \bm \Lambda_{cs}}\alpha(\bm{x}; M)$.
  \STATE evaluate the selected ${\bm x}_i$ and obtain its performance $y_i$.
  \STATE augment $D=D\cup({\bm x}_i,y_i)$.
  \ENDWHILE
  \STATE \textbf{return} the best configuration in $D$.

\end{algorithmic}
\end{algorithm}

Bayesian optimization (BO)~\cite{shahriari2015taking,hutter2011sequential,bergstra2011algorithms,li2021openbox} is a framework to solve black-box problems where the objective can only be observed via evaluation. 
The main advantages of BO are that it is robust to noise and can estimate uncertainty to balance exploration and exploitation.
A typical loop in vanilla BO contains the following four steps:
(1) it fits a \textit{surrogate model} $M$ based on observed configurations $D = \{(\bm{x}_1, y_1),...,(\bm{x}_{n-1}, y_{n-1})\}$, where $\bm{x}_i$ is the $i^{th}$  configuration and $y_i$ is the corresponding performance;
(2) it chooses the next promising configuration that maximizes the \textit{acquisition function} $\bm{x}_{n}=\arg\max_{\bm{x} \in \mathcal{X}}\alpha(\bm{x}; M)$, where the acquisition function is designed to balance the exploration and exploitation; (3) it evaluates the chosen configuration $\bm{x}_{n}$ to obtain its performance $y_{n}=f(\bm{x}_{n})+\epsilon$ with noise $\epsilon \sim \mathcal{N}(0, \sigma^2)$; 
(4) add the pair $(\bm{x}_{n}, y_{n})$ to observations so that $D = \{(\bm{x}_1, y_1),...,(\bm{x}_{n}, y_{n})\}$.
The pseudo-code is provided in Algorithm~\ref{algo:bo}.

\noindent
{\bf Surrogate. }
To model the objective function, different from machine learning algorithms used in offline approaches~\cite{bei2015rfhoc,yu2018datasize}, which incorporate additional algorithm hyperparameters, 
we choose Gaussian process (GP)~\cite{rasmussen2003gaussian} as the surrogate model because it is hyperparameter-free and provides closed-form inference.
GP is a flexible class of models for specifying prior distributions over objective function: $f: \mathcal{X} \rightarrow R$.
The convenient properties of the Gaussian distribution allow us to compute marginal and conditional means and variances in closed form. 
Given an unseen configuration $\bm x$, the predictive mean and covariance are expressed as follows,
\begin{equation}
\small
\begin{aligned}
    \mu(\bm x)&=K(\bm X,\bm x)(K(\bm X, \bm X)+\tau^2I)^{-1}\bm Y,\\
    \sigma^2(\bm x)&=K(\bm x, \bm x)+\tau^2I-(K(\bm X, \bm X)+\tau^2I)^{-1}K(\bm X, \bm x),
\end{aligned}
\end{equation}
where $K$ is the covariance matrix, $\bm X$ are the observed configuration vectors and $\tau^2$ is the level of white noise. 

\noindent
{\bf Acquisition Function. }
To estimate the performance improvement of the unseen configuration, we apply the Expected Improvement (EI)~\cite{jones1998efficient} function. 
Given the marginal predictive mean $\mu(\bm x)$ and variance $\sigma^2(\bm x)$ by the surrogate model, the EI function is defined as the expected improvement over the best performance found,
\begin{equation}
\small
\begin{aligned}
    EI(\bm x)&= \int_{-\infty}^{\infty} \max(y^*-y, 0)p_{M}(y|\bm x)dy\\
    &=\sigma(\bm x)\left(\gamma(\bm x)\Phi(\gamma (\bm x))+\phi(\gamma(\bm x))\right),
\end{aligned}
\label{eq:ei}
\end{equation}
where $\gamma(\bm x)=\frac{y^*-\mu(\bm x)}{\sigma(\bm x)}$, $y^*$ is the best performance observed so far. The second line of Equation~\ref{eq:ei} is the closed form of EI under Gaussian Process, $\Phi(\cdot)$ and $\phi(\cdot)$ are standard normal cumulative distribution function and probability density function, respectively.
While vanilla EI is designed for problems without constraints, we will introduce how we deal with constraints in Section~\ref{sec:eic}.

\noindent
{\bf Dynamic Workload Support. }
As aforementioned, the workload may change during the online tuning process, particularly the data size.
Since the same configuration for a workload may achieve different results with different input data sizes, the change in data size could affect the tuning result. 
To accommodate this, we take the data size along with configuration into consideration and model the objective value with the Gaussian Process. 
Concretely, the configuration ${\bm x}_i$ contains the configuration values and the dataset size, which is represented as ${\bar{\bm x}}_i=\{x_i^1,..., x_i^N, ds_i\}$, where $x_i^j$ is the value for the $i^{th}$ configuration and $ds_i$ is the data size of that run. 
After evaluating $n$ Spark configurations, we elaborate the objective $f( \bar{X})$ by Gaussian distribution as follows,
\begin{equation}
\small
    f(\bar{X}) \sim GP(0,K(\bar{X},\bar{X}')),
\end{equation}
where $K$ denotes the covariance matrix, and $\bar{X}$ is a $n*(N+1)$ matrix where $N$ is the number of Spark parameters.
$\bar{X}$ and $\bar{X}'$ are matrices of the same size.
We propose to use mixed kernels to implement this. 
Concretely, we apply the Matern kernel for numerical parameters, the Hamming kernel for categorical parameters, and the SE kernel to deal with data size.
However, due to data privacy issue, input data size is not always accessible in production tasks. 
In this case, besides Spark configurations, we train a datasize-aware surrogate by also taking the hour of the day or the current day of the week as input to characterize the periodic change of data.
Therefore, the BO framework can adapt to dynamic workloads in online tuning.

\noindent
{\bf Initial configurations.}
To estimate the performance in a wide area of configuration space, we sample several configurations using low-discrepancy
sequences~\cite{sobol1998quasi} to initialize the observations $D$ (Line 1 in Algo.~\ref{algo:bo}). 
In addition, to accelerate the convergence in the beginning, the meta-learning module can suggest the initial configurations based on task similarity, instead of using the low-discrepancy sequences.
We will describe the detail in Section~\ref{sec:initial-design}.

\noindent
{\bf Stopping \& Restarting Criterion. }
When the tuning budget exhausts or the stopping criterion is met, we stop the current tuning process and return the best configuration found so far for each upcoming config request.
When the expected improvement in Eq.~\ref{eq:ei} is less than a threshold (e.g. 10\%), the stopping criterion is activated. This prevents our framework from struggling to make small improvements.
In addition, the controller compares the difference between the expected performance and the actual execution result of the workload over runs.
If a continuous degradation is detected, re-tuning becomes necessary, and our framework will restart the tuning process, where meta-learning is utilized to extract knowledge from the previous runhistory to speed up the optimization.

\section{Efficient \& Safe Config Acquisition}
\label{sec:config_acquisition}
In this section, we present the methods in configuration generator for dealing with the slow convergence and safe exploration.

\subsection{Sub-space Generation}
As aforementioned, the complete configuration space is huge, which greatly limits the efficiency of the configuration search.
An intuitive idea is to use a smaller configuration sub-space that includes the most influential parameters, instead of the original huge space.
Tuning the sensitive parameters could bring significant performance improvement, and avoid wasting efforts in tuning those parameters that have little influence on performance.
Furthermore, the size of the sub-space is crucial: a large sub-space leads to slow convergence of configuration search but has a high probability of finding the optimal configuration;
BO tuning framework could converge to a local configuration quickly over a small sub-space, but this sub-space may not include the optimal or near-optimal configurations.
Therefore, when dealing with the high-dimensional space, we need to answer two questions: (1) how to measure the importance of Spark parameters, and (2) how to decide/adjust the size of the sub-space adaptively to pursue efficiency and effectiveness simultaneously.

\noindent
{\bf Parameter Importance \& Sub-space. }
To obtain the sub-space, we should first measure the importance of each parameter. 
While existing methods only consider the influence between each parameter and performance via Spearman Correlation Coefficient or weights of a learning model~\cite{fekry2020tuneful,locat}, we consider the importance of both single parameters and of interactions between parameters.
To this end, we adopt FANOVA~\cite{hutter2014efficient} to assess the importance of parameters.
It is a linear-time algorithm for computing marginals of random forest predictions and then leverages these predictions within a functional ANOVA framework to quantify the importance of both single parameters and of interactions between parameters.
Given the tuning history from a task, our parameter importance analysis module could rank the parameters according to their importance.

During optimization, we can get the importance score of parameters based on its tuning history for each task and obtain the final importance scores by averaging the scores from those tasks.
Based on this result, we define the sub-space as:
\begin{equation}
    \small
    \bm{\Lambda}_{sub}=\Lambda^1 \times ... \times \Lambda^K,
\end{equation}
where $\Lambda^i$ is the domain of parameter $x^i$ with the $i^{th}$ largest importance score, and $K$ is the size of the sub-space.
Note that, when starting tuning from zero, there is no tuning history available to obtain the parameter ranking based on the importance scores.
We start with an initial parameter ranking suggested by experts. 
Once new tuning history arrives, we will continuously update the importance score for each parameter based on FANOVA.

\noindent
{\bf Evolution Strategy of Sub-space. }
Different from existing methods that use a fixed sub-space~\cite{locat}, we propose to automatically adjust the size of the sub-space, i.e., $K$.
At the beginning of the tuning process, we initialize the size of sub-space $K$ to a small constant $K \leftarrow K_{init}$, and then the value of $K$ will be updated in a certain period, i.e., $N_{space}$ iterations.
The range of space size is $[K_{min}, K_{max}]$, where $K_{max}$ is the number of Spark parameters in the entire space $\bm \Lambda_{cs}$.
On one hand, a sub-space should be sufficiently large to contain good configurations.
On the other hand, it should be small
enough to ensure that BO is accurate and efficient within this sub-space. 
Therefore, the evolution of $K$ is crucial.

To dynamically control the space size, similar to TuRBO~\cite{eriksson2019scalable}, we adopt an intuitive design that we expand the sub-space when the BO optimizer ``makes progress'', i.e., it finds better configurations in the current sub-space, and shrinks it when the optimizer appears stuck.
We define a ``success'' as a configuration that improves over the best configuration found, and a ``failure'' if a configuration that fails.
After $\tau_{sucs}$ consecutive successes, we increase the size of the sub-space to $K \leftarrow \operatorname{min}(K_{max}, K + 2)$.
After $\tau_{fail}$ consecutive failures, we decrease the size of the sub-space to $K \leftarrow \operatorname{max}(K_{min}, K - 2)$.
Once the size of sub-space is changed, the counters for the success and failure events are set to zero.
Based on this mechanism, the sub-space can be adjusted adaptively to achieve high efficiency while getting good convergence results. In our implementation, $\tau_{sucs}=3$, $\tau_{fail}=5$, $K_{min}=4$ and $K_{init} = 10$, respectively.

\subsection{Safe Exploration and Exploitation}
\label{sec:eic}
To deal with constraints as we introduced in Eq.~\ref{eq:def:gentuning}, we apply the EI with constraints (EIC)~\cite{gardner2014bayesian} as the acquisition function. 
Different from the traditional EI~\cite{jones1998efficient} in Eq.~\ref{eq:ei}, EIC takes the probability of satisfying the constraints into consideration. 
Concretely, EIC for the generalized tuning in Eq.~\ref{eq:def:gentuning} is defined as follows, 
\begin{equation}
\small
    EIC(\bm x)=Pr[T(\bm x)\leq T_{max}] \cdot EI(\bm x),
\label{eq:eic}
\end{equation}
where $Pr[T(\bm x)\leq T_{max}]$ refers to the probability that satisfies the runtime constraint. 
To predict the runtime of an unseen configuration, we also fit a surrogate (i.e., GP) on runtime metrics. 
To customize other constraints, e.g., restricting resources, we can implement it by multiplying the term in Eq.~\ref{eq:eic} by $Pr[R(\bm x)\leq R_{max}]$.
Without loss of generality, we take the runtime constraint as an example. The probability can be computed as,
\begin{equation}
\small
    Pr[T(\bm x)\leq T_{max}]=\int_{-\infty}^{T_{max}}p(T(\bm x)|\bm x) dT(\bm x),
\end{equation}
which can be obtained by using the posterior prediction from GP.

In online tuning scenarios, evaluating a configuration that violates the constraints will inevitably downgrade the job performance. 
To explicitly ensure the safety of job execution, we further propose safe exploration and exploitation to select the next candidate.
Given the predictive mean $\mu_t^T(\bm x)$ and variance ${\sigma_t^T}^2(\bm x)$ from the runtime surrogate built at the $t^{th}$ iteration, we define the upper bound as:
\begin{equation}
\small
u_t^T(\bm x)=\mu_t^T(\bm x)+\gamma\sigma_t^T(\bm x),    
\end{equation}
where $\gamma$ is a constant that controls the bound and $\gamma \in (0, 1]$.
We maintain a safe region so that the configurations in that region are expected to satisfy the constraints in the worst case, which is theoretically analyzed in previous work ~\cite{sui2015safe}.
The safe region for runtime at the $t^{th}$ iteration is defined as
$S_t^T=\left\{ {\bm x} \in \bm \Lambda_{sub}\  |\  u_t^T(\bm x) \leq T_{max}  \right\}$. For multiple constraints, the final safe region is the intersection of single safe regions. 
The main process for the configuration generator to produce a new configuration is: (1) build the final safe region $S$, and (2) choose the next configuration by solving the EIC over $S$.

\subsection{Approximate Gradient Descent}
In Spark tuning, the objective in Eq.~\ref{eq:def:gentuning} is treated as a black-box function of input parameters with no available derivative information. 
However, some parts of the objective functions are not always black-box. 
Though the relationship between runtime and Spark parameters is complicated, the resource function has an analytic form, e.g., $R(\bm{x})=\#cpu\_vcores(\bm{x}) + c\cdot\#mem(\bm{x})$, which is directly determined by the values of parameters such as {\tt spark.executor.instances}, {\tt spark.executor.cores}, and {\tt spark.executor.memory}.
Gradient-based techniques~\cite{ruder2016overview} perform well in optimizing differentiable functions.
Compared with BO, the derivative information of the objective function provides more precise guidance on how to choose the next input parameters based on the current configurations (i.e., exploitation signal).
When observations $D$ are sufficient to approximate the objective function $f(\bm{x})$, we can estimate the derivative information.
However, since AGD emphasizes the exploitation signal only during the search process, it could be stuck in  local configurations easily.
To address this and integrate the benefits of both methods, we {\bf alternately apply AGD and BO} to select the next configuration so that the combined framework balances exploration and exploitation well with a convergence guarantee provided in previous work~\cite{hutter2011sequential}.
Therefore, we develop the approximate gradient descent (AGD) technique within BO to utilize the derivative information of the objective function and further accelerate the convergence over the search space.

To apply gradient descent, the partial derivative of $f(\bm x)$ over each numerical parameter $x^i$ is:
\begin{equation}
\small
    \frac{\partial f}{\partial x^i}(\bm x) = \beta (\frac{T(\bm x)}{R(\bm x)})^{\beta-1} \frac{\partial T}{\partial x^i}(\bm x) + (1-\beta) (\frac{T(\bm x)}{R(\bm x)})^{\beta} \frac{\partial R}{\partial x^i}(\bm x),
\end{equation}
where $x^i$ is the $i^{th}$ Spark parameter. Since the resource function $R(\bm x)$ is a white-box function, the partial derivative of $R(\bm x)$, i.e. $\partial R(\bm x) / \partial x^i$, has an analytical form. 
However, the partial derivative of the runtime function $T(\bm x)$, i.e. $\partial T(\bm x) / \partial x^i$, cannot be calculated directly. 
We approximate the partial derivative $\partial T(\bm x) / \partial x^i$ by:
\begin{equation}
\small
    \frac{\partial T}{\partial x^i}(\bm x) \approx \frac{T(\bm{x} + \bm{\epsilon_i}) - T(\bm{x} - \bm{\epsilon_i})}{2|\bm{\epsilon_i}|},
\end{equation}
where all dimensions except the $i^{th}$ element in  $\bm \epsilon_i$ are zero. The values of $T(\bm{x} + \bm{\epsilon_i})$ and $T(\bm{x} - \bm{\epsilon_i})$ are predicted by the surrogate for runtime. 
Then, the value of each parameter $x^i$ is updated by:
\begin{equation}
\small
    x^i \leftarrow x^i - \eta * \frac{\partial f}{\partial x^i}(\bm x),
\end{equation}
where $\eta$ is the learning rate with $\eta=0.001$. 
During the tuning procedure, AGD is integrated into BO. 
Every $N_{AGD}=5$ BO iterations, instead of BO, the next configuration $\bm{x}$ is generated by conducting AGD based on the best configuration found so far. 

\noindent
{\bf Algorithm Summary in Generator. }
Algorithm~\ref{algo:framework} gives the pseudo-code in configuration generation. 
During each iteration of BO, we train multiple surrogates for objective, runtime, other constraints and resource on current observations (Line 1). 
Every $N_{AGD}$ iterations, we choose the next configuration via AGD (Line 4).
Else, based on the constraint surrogates, we obtain the candidate region by intersecting the sub-space and safe regions for each constraint (Lines 6-7). 
Then, we choose the configuration that maximizes the EIC acquisition function over the candidate region (Line 8).
Finally, the suggested configuration is returned (Line 10).

\section{Meta-learning based Acceleration}
\label{sec:meta-learning}
In this section, we describe the design of the meta-knowledge learner, which utilizes meta-learning based techniques to accelerate the configuration search in BO.
We first introduce the method to learn the similarity between tuning tasks using the meta-features from workloads.
Then, based on similarity learning, we present the meta-learning methods for warm-starting and surrogate modeling.

\subsection{Task Characterization \& Similarity Learning}
Inspired by previous work~\cite{prats2020you}, we extract the feature vector of a tuning task (i.e., meta-features) from \texttt{SparkEventLog}, in which Spark event information is logged during execution.
The meta-features summarize information in two levels: stage and task.
Stage information contains the Spark actions and transformations used in the Stage, which shows core Spark function calls made during the job execution.
Task information describes whether, overall, the task was read or write-intensive, CPU intensive, etc. 
The resulting meta-features include a total of 75 features, where 11 are retrieved from Stage information and 64 are retrieved from Task information~\cite{prats2020you}.

A straightforward method that measures the similarity is to compute the Euclidean distance between the two meta-features of corresponding tasks.
However, the type and scale of each meta-feature are heterogeneous, which greatly decreases the effectiveness of Euclidean distance.
Besides, each meta-feature poses a diverse impact on similarity learning.
To address these issues, we propose to use a supervised learning method (i.e., regression model) to learn the similarity given two tasks.
Concretely, given the meta-features of two input tasks: $\bm{v_1}$ and $\bm{v_2}$, the regression model $\mathcal{M}_{reg}: (\bm{v_1}, \bm{v_2}) \mapsto d$ predicts their distance $d \in [0, 1]$.
A smaller distance $d$ indicates that the two tasks are more similar to each other.

\begin{algorithm}[t]
  \small
  \caption{Pseudo code for configuration generation.}
  \label{algo:framework}
  \begin{algorithmic}[1]
  \REQUIRE Evaluation observations $D$, constants, e.g., $N_{AGD}$, $\eta$, etc.
  \ENSURE the suggested configuration $\bm{x}_i$ for the $i^{th}$ iteration.
  \STATE Trains surrogates (e.g., $M^f$, $M^T$, $M^R$) on current observations $D$.
  \IF{$|D|+1 \ \text{mod} \  N_{AGD}=0$}
  \STATE Set current best configuration: $\bm{x}_i \leftarrow \operatorname{get\_best\_config}(D)$. 
  \STATE Choose the next configuration via AGD: $x^i \leftarrow x^i - \eta * \frac{\partial f}{\partial x^i}(\bm x)$.
  \ELSE  
  \STATE Update the size of the sub-space $\bm \Lambda_{sub}$ based on runhistory $D$.
  \STATE Build the safe region for all constraints: $S_i = S^T_i \cap S^R_i \cap \bm\Lambda_{sub}$.
  \STATE Choose the configuration by solving $\bm{x}_{i}=\arg\max_{\bm{x} \in S_i}EIC(\bm{x})$.
  \ENDIF
  \STATE \textbf{return} the Spark configuration $\bm{x}_i$.
\end{algorithmic}
\end{algorithm}

To train such a regression model, we prepare the training data as follows.
We collect a wide range of Spark jobs $T_{1,...,K}$ and the corresponding tuning history $H_{1, ..., K}$, where the tuning history $H_{i}$ of the $i^{th}$ task consists of the pairs of configuration and its performance.
As aforementioned, the meta-feature for each tuning task can be obtained by parsing the information in event log~\cite{prats2020you}; next, we need to define the distance metric. 
In the context of Spark tuning, we focus on the partial orderings over configuration performance.
To this end, we measure the distance by calculating the number of discordant pairs of predictions (i.e., pair-wise ranking) using the \textbf{negative} Kendall-tau coefficient.
That is, given two configurations $\bm{x}_k$ and $\bm{x}_l$, and surrogates model $M^i$ and $M^j$ on the $i^{th}$ and $j^{th}$ task, the pair $(\bm{x}_k,\bm{x}_l)$ is discordant if the sort order of $(M^i(\bm{x}_k),M^i(\bm{x}_l))$ and $(M^j(\bm{x}_k),M^j(\bm{x}_l))$ disagrees.
Then, the distance is defined as \textbf{the ratio of discordant pairs}, which is $Dist(M^i,M^j)=\frac{1-\tau^{D_{rand}}(M^i,M^j)}{2}$ where $\tau^{D_{rand}}(M^i,M^j)$ is the Kendall-tau coefficient of two vectors by applying surrogates $M^i$ and $M^j$ on randomly sampled configurations $D_{rand}$. While the range of Kendall-tau coefficient is $[-1,1]$, we scale the range of distance $Dist(M^i, M^j)$ to $[0, 1]$.
Finally, we obtain the all basic ingredients to learn the meta-learner $\mathcal{M}_{reg}$ that could predict the distance between two tasks.
In our implementation, we train a LightGBM~\cite{ke2017lightgbm} regressor to obtain $\mathcal{M}_{reg}$.


\subsection{Warm-starting \& Surrogate Improvement}
Based on similarity learning, we could speed up the configuration search in two aspects: initial design and surrogate modeling.

\noindent
{\bf Initial design with warm-starting. }
\label{sec:initial-design}
In vanilla BO, the search process starts from scratch. That is, it initializes observations $D$ with several evaluations of random configurations (in Algorithm~\ref{algo:bo}).
By leveraging similarity across tasks, given a new tuning task, we can obtain similarity results between the new task and previous tasks. 
We rank the previous tasks using the predictions of meta-learner $\mathcal{M}_{reg}$, and choose the top-3 most similar tasks. 
Then, we select the best Spark configuration found in these top-3 tasks, and set them as the initial configurations before starting the main BO loop in Algorithm~\ref{algo:bo} (Line 1).
In this way, our framework could achieve a better performance in the beginning.
In addition, we also can suggest the sub-space for a new tuning task using task similarity.

\noindent
{\bf Surrogate modeling with meta-learning. }
Given scarce observations at the beginning of the optimization, BO can not build an accurate surrogate to guide the search.
In this case, BO fails to converge to optimal configuration quickly.
To overcome this issue, we leverage the tuning history from previous similar tasks to obtain an accurate surrogate. 
Therefore, we propose to build a meta-learning surrogate ensemble $M_{\text{meta}}$.
The prediction of $M_{\text{meta}}$ at configuration $\bm{x}$ is given by $y \sim \mathcal{N}(\mu_{\text{meta}}(\bm{x}), \sigma^2_{\text{meta}}(\bm{x}))$:
\begin{equation}
\small
\begin{aligned}
\mu_{meta}(\bm{x})=\sum_iw_i\mu_{i}(\bm{x}), \quad \sigma^2_{meta}(\bm{x})=\sum_iw_i^2\sigma_i^2(\bm{x}),
\end{aligned}
\end{equation}
where 
$\mu_i$ and $\sigma^2_i$ are the predictive mean and variance from base surrogate $M^i$.
The weight of base surrogate ${w}_i$ reflects the similarity between the previous and current task. Here we set the weight to ${w}_i = 1 - Dist(M^i,M^t)$ and then normalize the weights with $\sum_iw_i=1$.
In addition, $M_{meta}$ also includes a surrogate $M^t$ trained on the current task, and the weight of $M^t$ is determined by cross-validation strategy~\cite{feurer2018scalable}.
By integrating useful information from both the current task and previous tasks, $M^{\text{meta}}$ could offer more accurate modeling for the objective function, thus further accelerating the convergence of the current tuning task.


\section{Experiments and Results}
To evaluate our framework, in this section, we list three insights we want to investigate: 
1) \textbf{Practicality} - Our framework achieves significant cost reductions compared with human experts when tuning 25K real-world Spark tasks in Tencent.
2) \textbf{Generality} - Our framework supports different tuning objectives (runtime and cost) and Spark task types (MR and SQL), and consistently outperforms state-of-the-art Spark tuning approaches on various standard benchmarks; 
3) \textbf{Efficiency} - Our framework equipped with the three techniques greatly accelerates the tuning process and requires only a few trials to find near-optimal configuration.

\subsection{Experiment Setup}

\noindent
\textbf{Benchmark Programs:}
In addition to the real-world production tasks (described in Section~\ref{sec:exp_production}), we also use programs from the well-known big data benchmark suite HiBench~\cite{huang2010hibench} in our experiments.
HiBench contains various Spark tasks, including machine learning, graph, websearch, etc.
We choose 6 representative tasks: Bayes, KMeans, NWeight, WordCount, PageRank, and TeraSort. A larger set with 16 tasks is used in the meta-learning experiment.

\noindent
\textbf{Environment:}
When tuning the in-production Spark tasks, the online job executions are conducted on the resource group of our customers, which is allocated by the Tencent data platform.
For example, one of our customers owns a resource group with 100 computing units, where each unit corresponds to 20 Intel(R) Xeon(R) Platinum 8255C CPU cores with a base frequency of 2.50GHz and 50 GB memory.
The HiBench tasks are executed on a small x86 cluster with four nodes. 
Each server is equipped with 2 AMD EPYC 7K62 2.80GHz 48-core processors and 512GB PC4 memory. 
In this environment, we use Spark 3.0 as our computing framework.

\noindent
\textbf{Spark Parameters:}
We use the same Spark configuration parameters as used in Tuneful~\cite{fekry2020tuneful}, which contains 30 parameters that significantly influence the application performance.
The value ranges of the parameters are set differently depending on the cluster size.

\noindent
\textbf{Compared Methods.}
When tuning Hibench tasks, we compare our framework with the following approaches: (1) \textit{Random Search}~\cite{bergstra2012random} generates a random configuration at each iteration.
(2) \textit{RFHOC}~\cite{bei2015rfhoc} trains several random forests for each task, and utilizes the generated models along with a genetic algorithm to explore the configuration space.
(3) \textit{DAC}~\cite{yu2018datasize} is a datasize-aware auto-tuning approach, which utilizes hierarchical regression tree models and genetic algorithm to efficiently identify the near-optimal configurations. 
(4) \textit{CherryPick}~\cite{alipourfard2017cherrypick} uses Bayesian optimization (BO) to find the optimal configuration for Spark applications. The model aims at minimizing the user cost and subjects to a runtime threshold.
(5) \textit{Tuneful}~\cite{fekry2020tuneful} utilizes BO to tune the configuration of in-memory cluster computing systems in an online manner.
(6) \textit{LOCAT}~\cite{locat} is also a BO-based online approach to conduct auto-tuning for Spark SQL.

\noindent
\textbf{Objectives.}
To evaluate the generality, we consider two tuning objectives in the experiments:
(1) The \textit{Runtime} of Spark job, where $\beta=1$ in Eq.~\ref{eq:def:gentuning}.
(2) The execution \textit{Cost} of Spark job, where $\beta=0.5$ in Eq.~\ref{eq:def:gentuning}.
Note that the scale of the cost depends on the definition of runtime (e.g., in seconds or hours) and resources in different experiments.
The tuning objective of RFHOC, DAC, Tuneful, and LOCAT, is runtime.
In the cost experiment, we modify some modules in their implementations to support cost minimization. 

\begin{figure*}[tb]
	\centering
 	\subfigure[Memory usage reduction]{
            \label{fig:memory_reduction}
		\scalebox{0.3}[0.3]{
			\includegraphics[width=1\linewidth]{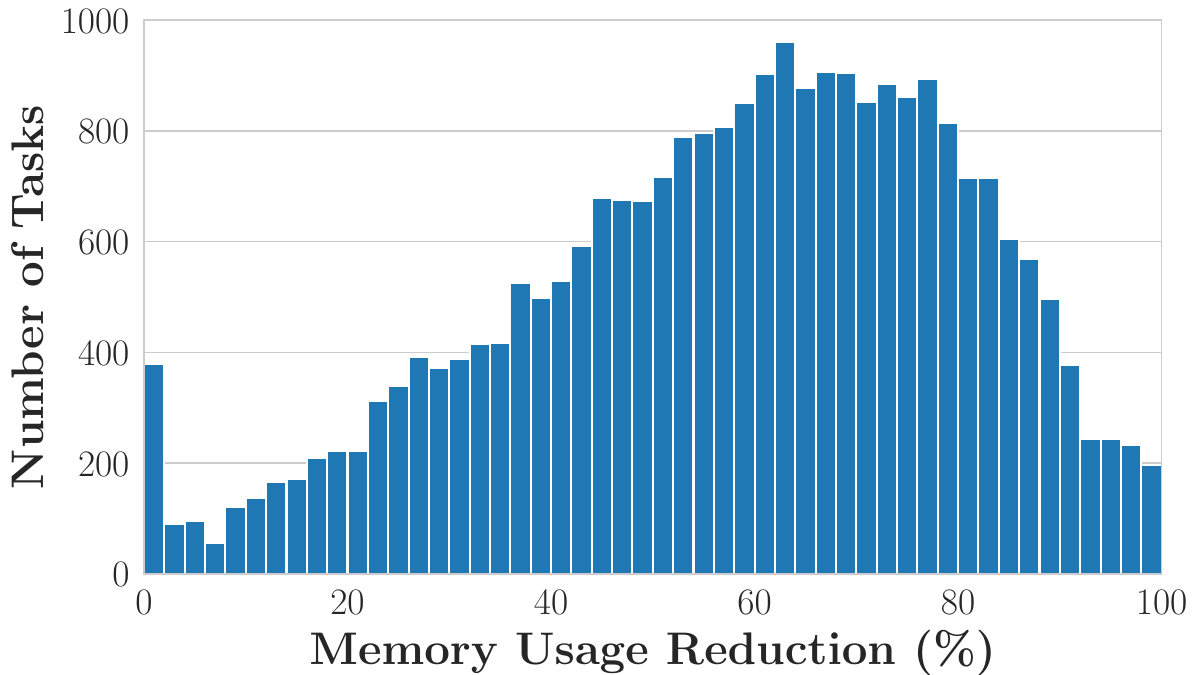}
	}}
	\subfigure[CPU usage reduction]{
            \label{fig:cpu_reduction}
		\scalebox{0.3}[0.3]{
			\includegraphics[width=1\linewidth]{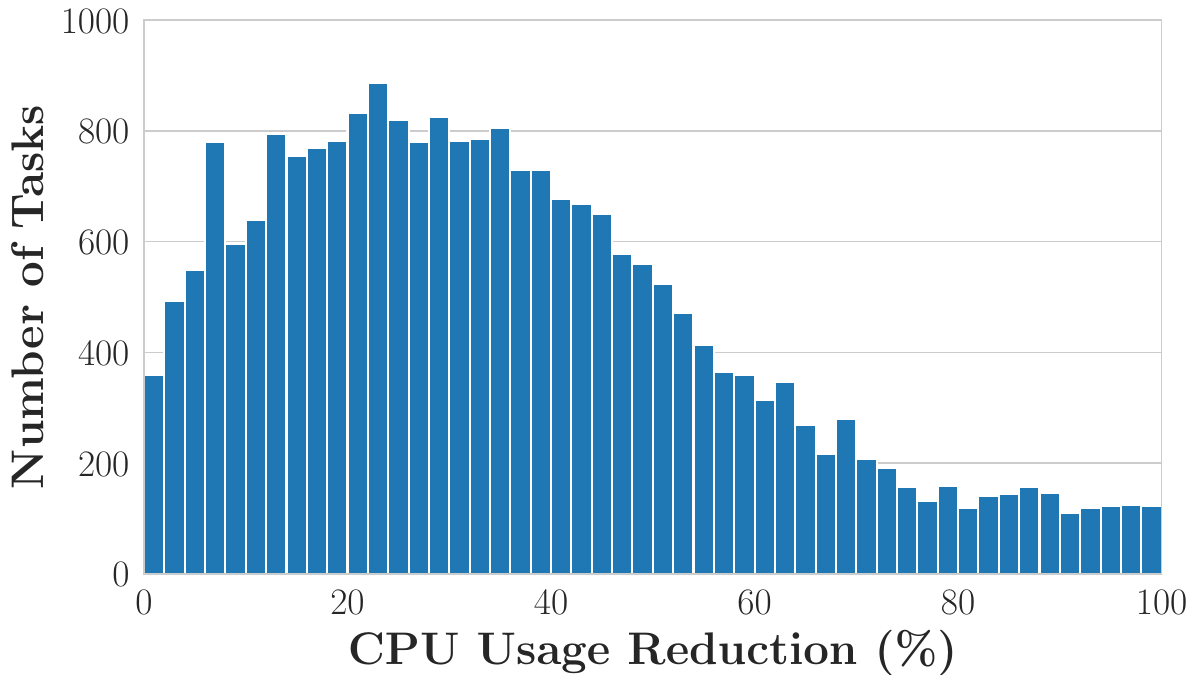}
	}}
	\subfigure[Average execution cost reduction]{
            \label{fig:convergence}
		\scalebox{0.3}[0.3]{
			\includegraphics[width=1\linewidth]{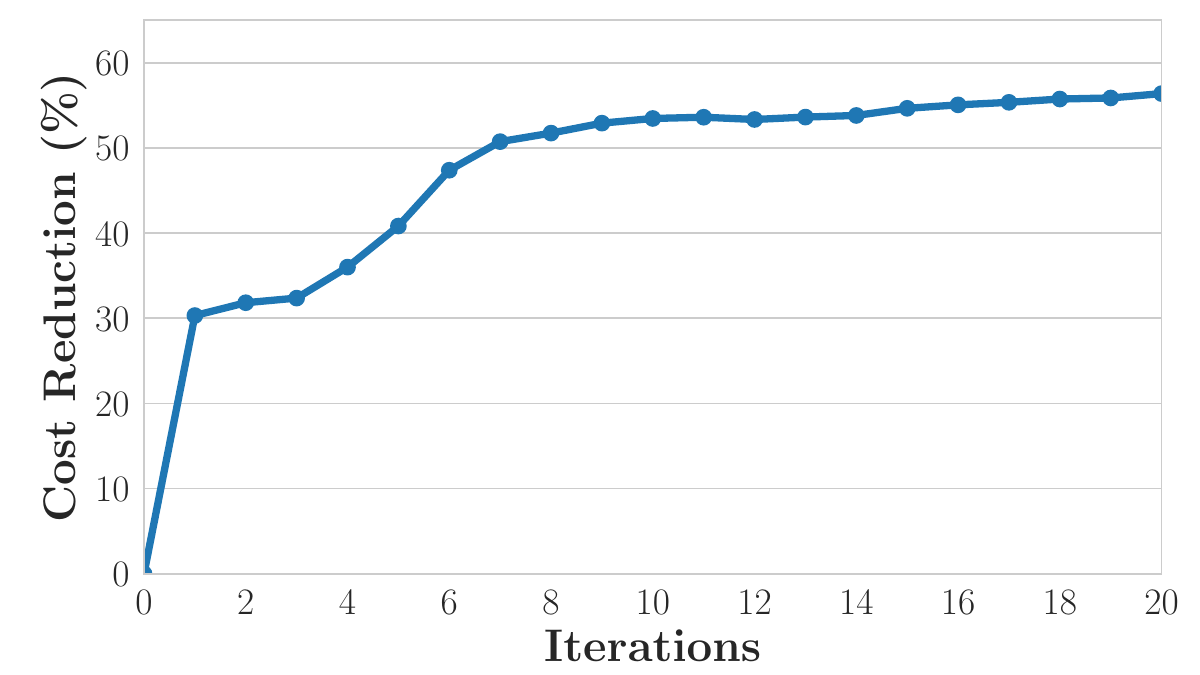}
	}}
	\caption{The reduction results of memory usage, CPU usage, and objective value on 25K Spark tasks in Tencent.}
\end{figure*}

\noindent
\textbf{Metrics \& Settings. }
We use the following three metrics:
(1) The \textit{Speedup} of execution time of the best-found configuration relative to the one from random search.
(2) The \textit{Cost} at the $i^{th}$ iteration, i.e., the execution cost of the $i^{th}$ tried configuration as defined in Eq.~\ref{eq:def:gentuning}. 
Note that, the \textit{\bf Min Cost} refers to the execution cost of the best configuration found so far.
(3) The \textit{Cost Reduction} relative to a reference method (e.g., random search or default configuration), which is defined as $\frac{Cost_{ref}-Cost}{Cost_{ref}}$.
The hyperparameters in our framework mentioned in previous sections are obtained via sensitivity analysis, and we apply the same settings in our implementation.
The BO component is implemented using the OpenBox library~\cite{jiang2023openbox}.
All experiments running on HiBench are repeated 10 times with different random seeds and the average metrics are reported.

\begin{figure}[tb]
    \centering
    \scalebox{0.95}[.95] {
        \includegraphics[width=1\linewidth]{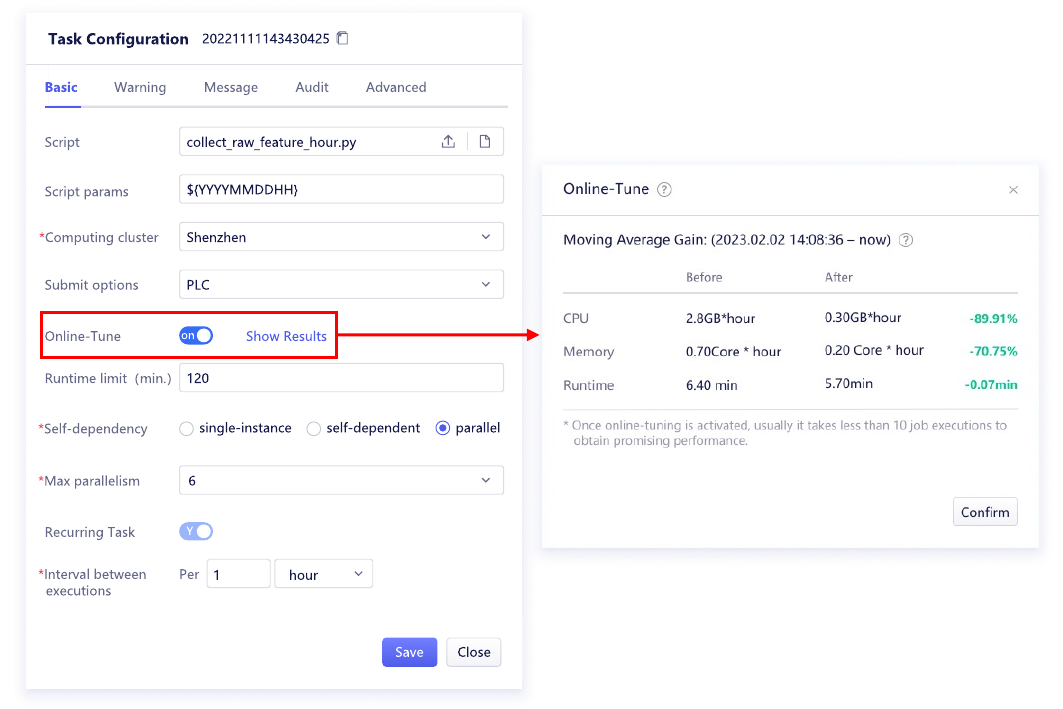}
    }
    \caption{Web UIs of Online-Tune provided by Tencent's internal data platform.}
    \label{fig:web_ui}
\end{figure}

\begin{table*}[htb]
\centering
\small
\caption{Detailed comparison between manual and tuned configurations on eight in-production tasks related to advertisement.}
\resizebox{1.98\columnwidth}{!}{
\begin{tabular}{ll|cccccccc}
\toprule
Task  & Method  &  Memory\_usage & CPU\_usage & Runtime(s) & Execution cost & Executor.instances & Executor.cores & Executor.memory(GB) & \#Iteration \\
\hline

\multirow{2}{*}{Spark: Feature Extraction} &
      Manual & 1776.86 & 834.18 & 6768.99 & 208125.83 & 300 & 2 & 8 & -
 \\
      & Ours & 811.12 & 273.82 & 5316.64 & 55718.35 & 183 & 3 & 1 & 12 \\
\hline
\multirow{2}{*}{Spark: User-Traffic Distrib.} &
      Manual & 3865.39 & 966.35 & 5543.30 & 170015.02 & 256 & 2 & 8  & -\\
      & Ours & 1095 & 1117.16 & 4401.42 & 131699.25 & 300 & 3 & 3 & 14
\\
\hline
\multirow{2}{*}{Spark: DAU Analysis} &
      Manual & 2513.62 & 628.07 & 1220.81 & 67727.11 & 500 & 4 & 16  & -\\
      & Ours & 495.20 & 248.31 & 1270.04 & 28008.57 & 728 & 1 & 2  & 8\\

\hline
\multirow{2}{*}{Spark: Log Processing} &
      Manual & 2462.04 & 1076.11 & 4068.47 & 662950.82 & 656 & 4 & 9 & -\\
      & Ours & 885.90 & 444.18 & 2219.14 & 49205.71 & 732 & 1 & 2 & 9\\

\hline
\multirow{2}{*}{Spark SQL: Data Selection} &
      Manual & 0.6731 & 0.6770 & 48.29 & 243.53 & 16 & 6 & 6 & -\\
      & Ours & 0.0189 & 0.0135 & 32.26 & 30.78 & 1 & 1 & 1 & 10\\

\hline
\multirow{2}{*}{Spark SQL: Skew Detection} &
      Manual & 9.4702 & 0.9788 & 227.03 & 667.60 & 20 & 2 & 20 & -\\
      & Ours & 0.4812 & 0.2883 & 213.08 & 147.82 & 1 & 1 & 1 & 7 \\

\hline
\multirow{2}{*}{Spark SQL: Feature Calculation} &
      Manual & 1374.84 & 0.93 & 518.34 & 321.78 & 3 & 2 & 1 & -\\
      & Ours & 277.57 & 0.40 & 497.00 & 210.59 & 2 & 1 & 1 & 13\\

\hline
\multirow{2}{*}{Spark SQL: Data Preprossing} &
      Manual & 34.7431 & 0.0222 & 23.1300 & 210.52 & 3 & 2 & 6 & - \\
      & Ours & 10.4521 & 0.0102 & 19.9520 & 103.13 & 1 & 2 & 1 & 6\\

\hline
Avg Reduction on 8 tasks & - & {\bf-76.52\%} & {\bf-56.29\%} & {\bf-17.58\%} & {\bf-62.22\%} & - & - & - & {\bf 9.88}\\
\bottomrule
\end{tabular}
}
\label{tab:overall_reduction}
\end{table*}

\subsection{Results on In-production Spark Tasks}
\label{sec:exp_production}

\noindent
{\bf (Implementation \& Application in Tencent)} 
We have implemented the proposed framework as an independent tuning service deployed on Tencent Cloud, and this service has been applied to Tencent's internal data platform.
It continuously tunes the recurring Spark jobs, which are scheduled and executed every hour, day, etc.
The data platform also provides web UIs for users to start online tuning and observe the tuning results conveniently (See web pages in Figure~\ref{fig:web_ui}).
The tuning objective is set to execution cost with $\beta=0.5$ in Eq.~\ref{eq:def:gentuning}, and the constraints are set to twice the metrics of the manual configurations.
Our framework optimizes the tuning objective (i.e., execution cost), which is positively correlated with the actual cost (i.e., {\bf both} memory and CPU usage).
In order to measure the actual cost directly and quantitatively, we also analyze the performance of configurations using the two metrics: memory usage (GB$\cdot$hour), CPU usage (core$\cdot$hour).

\noindent
{\bf (Large-Scale Tuning Results)}
To demonstrate the practicability and efficiency of our service in real-world scenarios.
We utilize it to tune around 25K in-production Spark tasks from Tencent's business scenarios, including advertisement, marketing, and social networking. 
These tasks are executed every hour and produce essential statistics for downstream tasks.
Before auto-tuning, the configuration used for each Spark task is manually tuned by big data engineers (abbr. manual). 
After activating online tuning, each task is tuned once an hour, and the total budget is set to 20 hours (i.e., 20 trials or iterations).
Figures~\ref{fig:memory_reduction} and ~\ref{fig:cpu_reduction} show the number of tuning tasks with different cost reduction results after tuning. 
We observed that, the average memory and CPU cost reduction of the tuned configuration compared with manual configuration on 25K tasks are 57.00\% and 34.93\%, respectively.
Moreover, 66.49\% of the tasks get a memory usage reduction of over 50\%, and 64.70\% of the tasks achieve a CPU usage reduction of over 25\%.
In Figure~\ref{fig:convergence}, we plot the average execution cost reduction ratio of the best configuration found during optimization on the 25K tasks. 
The tuning objective (i.e., execution cost) reduces significantly during early iterations, where we observe a reduction of 52.44\% within only 9 iterations.
Note that, our warm-starting technique with meta-learning used in the first 3 iterations leads to a huge improvement. 
This demonstrates the high efficiency of our service.

\begin{table}[tb]
\centering
\small
\caption{Averaged cost reduction (\%) of job execution metrics of under-tuning (under) within 20 iterations, and post-tuning (post) compared with pre-tuning (pre) on 25K real-world tasks in Tencent. `-' indicates the increase in metrics.}
\resizebox{0.98\columnwidth}{!}{
\begin{tabular}{l|cc}
\toprule
Metric & Cost Reduction(under vs. pre) & Cost Reduction(post vs. pre)\\
\hline
Memory usage & 2.28\% & \textbf{57.00\%} \\
CPU usage & -5.82\% & \textbf{34.93\%}\\
Runtime & 1.63\% & \textbf{10.72\%} \\
\bottomrule
\end{tabular}}
\label{tab:cost}
\end{table}

\begin{figure*}[tb]
	\centering
		\scalebox{0.9}[.9] {
		\includegraphics[width=1\linewidth]{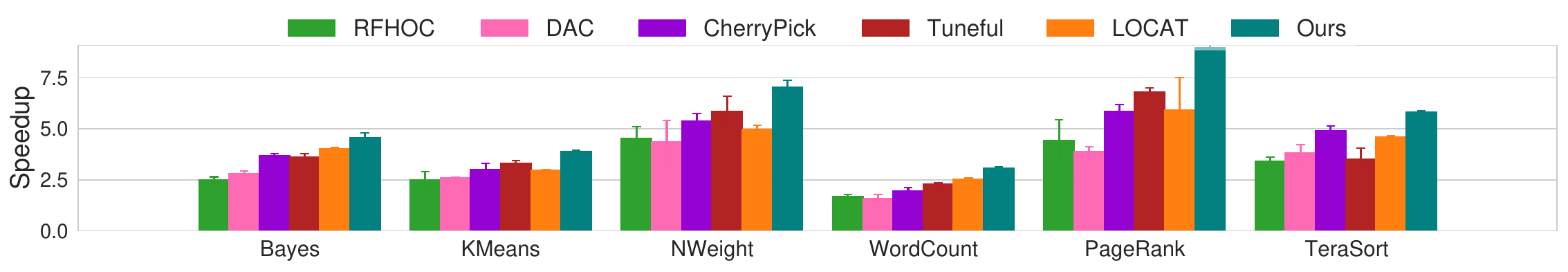}
         }
    \caption{Speedup of compared methods relative to random search on 6 HiBench tasks.}
    \label{fig:exp_main_runtime}
\end{figure*}
\begin{figure*}[tb]
	\centering
		\scalebox{0.9}[.9] {
		\includegraphics[width=1\linewidth]{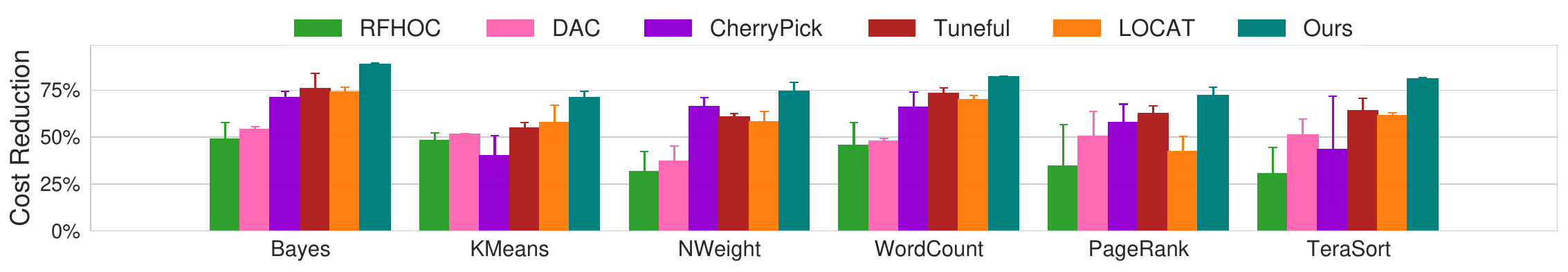}
         }
    \caption{Cost reduction of compared methods relative to random search on 6 HiBench tasks.}
    \label{fig:exp_main_cost}
\end{figure*}

\noindent
{\bf (Tuning Overhead Analysis)}
The tuning process is conducted only once, while the Spark job is repeatedly executed for a long time (e.g., months or years). 
The benefits of saving costs during each execution would accumulate, which is a large gain as time proceeds.
We also demonstrate the {\bf gains} and {\bf overheads} of online tuning on 25K tasks in Table~\ref{tab:cost}.
By comparing the post-tuning results (post) with manual pre-tuning results (pre), we observe significant gains on the three metrics (post vs. pre) for each job execution on average. 
For example, the tuned configuration reduces 57.0\% of the memory usage compared with the manual one.
In addition to the gains, we also compute the additional overhead (i.e., execution cost) by comparing the average metrics of each job execution during the tuning process (under) with those in pre-tuning. 
Since our framework needs to try configurations with different performance before finding a good one, the reduction ratios are lower during the tuning process (under vs. pre).
But after 20 iterations, the framework stops tuning and directly applies the best-found configuration so that the additional tuning overhead will be amortized quickly as the Spark tasks execute continuously.
Based on the results in Table~\ref{tab:cost}, our service needs {\bf no more than 4 extra executions to amortize the CPU usage overhead on average}.
Besides, the average metrics of the suggested configurations satisfy the constraints during tuning.

\noindent
{\bf (Analysis on Eight Specific Tasks)} 
To further demonstrate the generality and analyze the tuning behavior of our framework, we evaluate it on eight different online Spark tasks in detail and show the results in Table~\ref{tab:overall_reduction}. 
Each task is collected from the advertisement business in Tencent, where the first four traditional Spark tasks are executed once a day and the other four Spark SQL tasks are executed once an hour (generality on task types).
In the tuning process, all Spark tasks are tuned by trying different configurations in the production environment. 
We compare our framework with the manual configuration.
The tuning objective and constraints keep the same as in the previous 25K tasks. 
Compared with manually tuned configurations, our framework reduces the execution cost by 62.22\% on average. 
Meanwhile, we observe significant decreases \textbf{along with} the objective on the two metrics (i.e., average memory usage and CPU usage), which are 76.52\% and 56.29\%, respectively. 
Notably, we also report the number of iterations to achieve the tuned results, and the average iteration is {\bf less than 10}, which demonstrates the efficiency of our framework. 
In addition to the tuning results, we also provide the parameter values related to the Spark executor for reference. 
For example, in the first task -- feature extraction, compared with the manual configuration, spark.executor.instances (i.e., the number of executor instances) decreases from 300 to 183, spark.executor.cores increase from 2 to 3, and spark.executor.memory decreases from 8GB to 1GB.

\subsection{Results on Public Benchmarks}

To further investigate the generality, we conduct end-to-end experiments on 6 HiBench tasks on the small cluster. 
All methods are tested using two tuning objectives: (1) runtime with $\beta=1$, and (2) cost by setting $\beta=0.5$ in Eq.~\ref{eq:def:gentuning}.
The overall budget is set to 30 iterations.
We also set a runtime constraint in the experiment, where the threshold is twice the runtime of the default configurations.
This constraint ensures safety in online tuning so that the tuning procedure will not greatly affect online executions. 
For each method, we report the best observed objective value within 30 trials.

Figure~\ref{fig:exp_main_runtime} shows the speedup of all baselines relative to random search when the objective is runtime.
We observe that:
(1) The ML-based approaches, RFHOC and DAC, achieve a relatively lower speedup compared with Bayesian optimization-based methods.
ML models often need a large number of training samples, and 30 iterations are not sufficient to identify a promising configuration in the search space.
(2) The Bayesian optimization-based methods, CherryPick, Tuneful, LOCAT, and ours, get a better result under the limited budget. 
However, CherryPick does not reduce the dimension of search space when training the surrogate model, thus it cannot handle the large Spark search space well.
Tuneful and LOCAT use different techniques to select important parameters in the optimization process, but require 10 to 20 executions before shrinking the search space. Across the six tasks, their rankings are unstable.
(3) Our framework achieves the best and consistent speedup among all compared methods. 
Concretely, it leads to 3.08x-8.96x average speedups relative to random search, where the second best baselines in each task achieve only 2.54x-6.80x speedups.

In Figure ~\ref{fig:exp_main_cost}, we also present the relative execution cost reduction of all methods compared with random search. 
Compared with runtime, the execution cost is a more difficult objective to optimize. 
Concretely, our framework achieves a cost reduction of 71.22-88.97\% relative to random search and obtains a cost reduction of 38.43\% and 45.20\% on average compared with competitive baselines Tuneful and LOCAT, respectively.

\begin{table}[t]
\centering
\caption{Execution cost of the top-3 configurations found by our warm-starting method.}
\resizebox{.98\columnwidth}{!}{
\begin{tabular}{cc|ccccc}
\toprule
Target Task & Source Task & Default & Manual & Top1 & Top2 & Top3 \\
\hline
TeraSort & Sort & 844.70 & 91.3 & 54.51 & 40.66 & 43.77 \\
TeraSort & WordCount & 835.00 & 131.60 & 97.48 & 113.30 & 104.71 \\
LR & PageRank & 1431.21 & 245.90 & 183.35 & 333.39 & 214.73 \\
KMeans & SVD & 400.92 & 232.33 & 136.20 & 166.41 & 171.57 \\
\bottomrule
\end{tabular}}
\label{tab:exp:warm_start}
\end{table}

\begin{figure}[tb]
	\centering
	\subfigure[WordCount]{
		\scalebox{0.47}[0.47]{
			\includegraphics[width=1\linewidth]{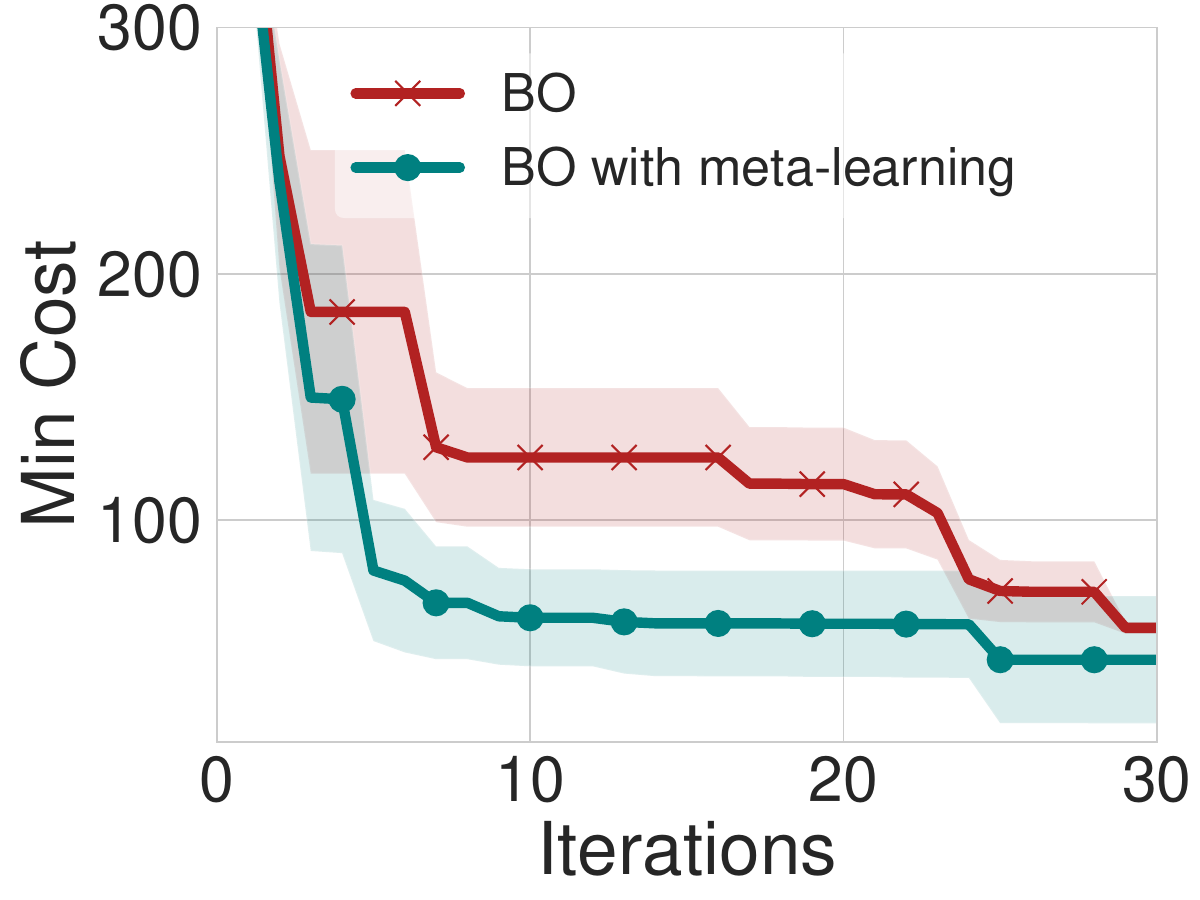}
	}}
	\subfigure[TeraSort]{
		\scalebox{0.47}[0.47]{
			\includegraphics[width=1\linewidth]{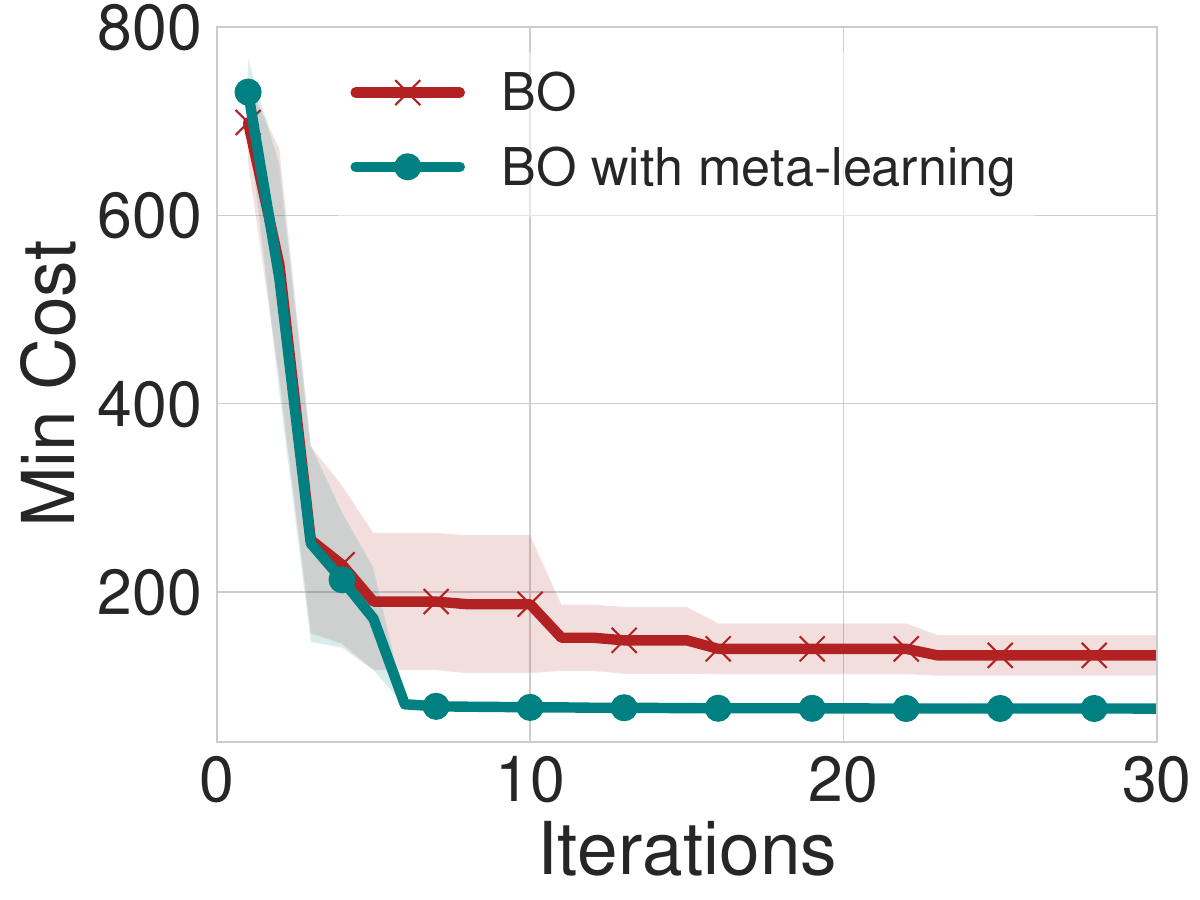}
	}}
	\caption{Results of tuning KMeans and TeraSort using BO with and without the meta-learning surrogate.}
 \label{fig:exp_metalearning_surrogate}
\end{figure}

\subsection{Meta-learning Experiments}
We conduct two experiments to evaluate the efficiency of meta-learning from the following two aspects:

\noindent
{\bf Warm-starting. }
Table~\ref{tab:exp:warm_start} shows the evaluation cost of configurations given by the warm-starting module in our meta-knowledge learner on three target tasks. 
The results of default and manually tuned configurations are slightly different on TeraSort (Lines 1-2) since we report the results of two independent evaluations on a computing cluster.
We observe that, by transferring the top-3 configurations from similar source tasks, the evaluation cost can be greatly reduced even in the initial 3 trials. 
Note that, the best configuration from similar source tasks can be a good one, but not always the best one.
The third-best configuration from Sort outperforms its best one on TeraSort, which implies the essence of warm-starting by transferring multiple good configurations instead of only one.
To summarize, the warm-starting module reduces the evaluation cost within three iterations by 66.03-95.19\% and 25.44-55.93\% relative to the default and manually-configured settings, respectively.

\begin{table}[tb]
\centering
\caption{Top-10 Spark parameters ordered by importance.}
\resizebox{.85\columnwidth}{!}{
\begin{tabular}{ccc}
\toprule
\# & Parameter Name               & Importance Score (mean $\pm$ std)     \\
\hline
1  & spark.executor.instances     & 0.3788 $\pm$ 0.1965 \\
2  & spark.executor.memory        & 0.1501 $\pm$ 0.1365 \\
3  & spark.memory.storageFraction & 0.0469 $\pm$ 0.0400 \\
4  & spark.default.parallelism    & 0.0366 $\pm$ 0.0530 \\
5  & spark.memory.fraction        & 0.0345 $\pm$ 0.0360 \\
6  & spark.executor.cores         & 0.0236 $\pm$ 0.0618 \\
7  & spark.io.compression.codec   & 0.0199 $\pm$ 0.0290 \\
8  & spark.shuffle.file.buffer    & 0.0146 $\pm$ 0.0187 \\
9  & spark.shuffle.compress       & 0.0138 $\pm$ 0.0142 \\
10 & spark.serializer             & 0.0083 $\pm$ 0.0099 \\
\bottomrule
\end{tabular}}
\label{tab:parameter_importance}
\end{table}

\begin{figure}[tb]
	\centering
	\subfigure[Tuning results]{
		\scalebox{0.47}[0.47]{
			\includegraphics[width=1\linewidth]{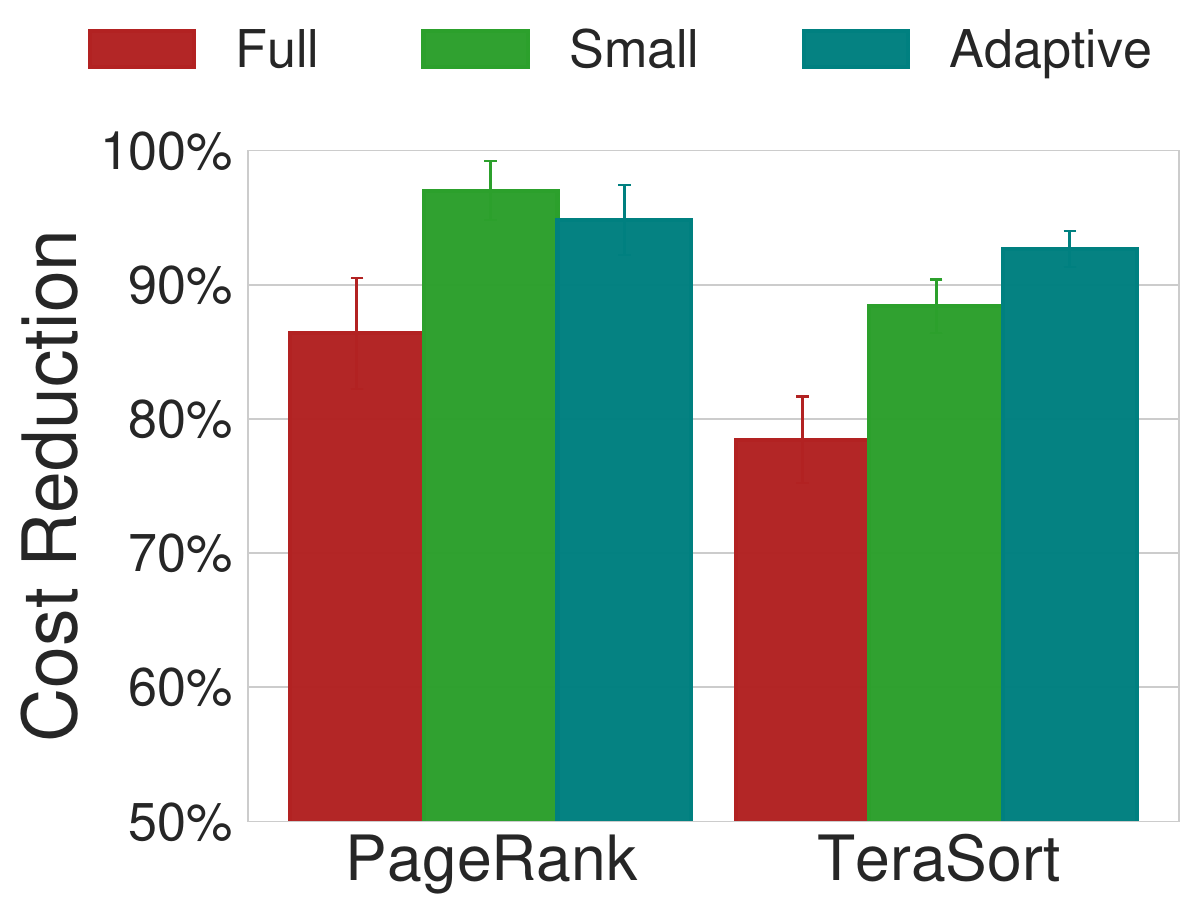}
            \label{fig:subspace:a}
	}}
	\subfigure[Optimization curve on Terasort]{
		\scalebox{0.47}[0.47]{
			\includegraphics[width=1\linewidth]{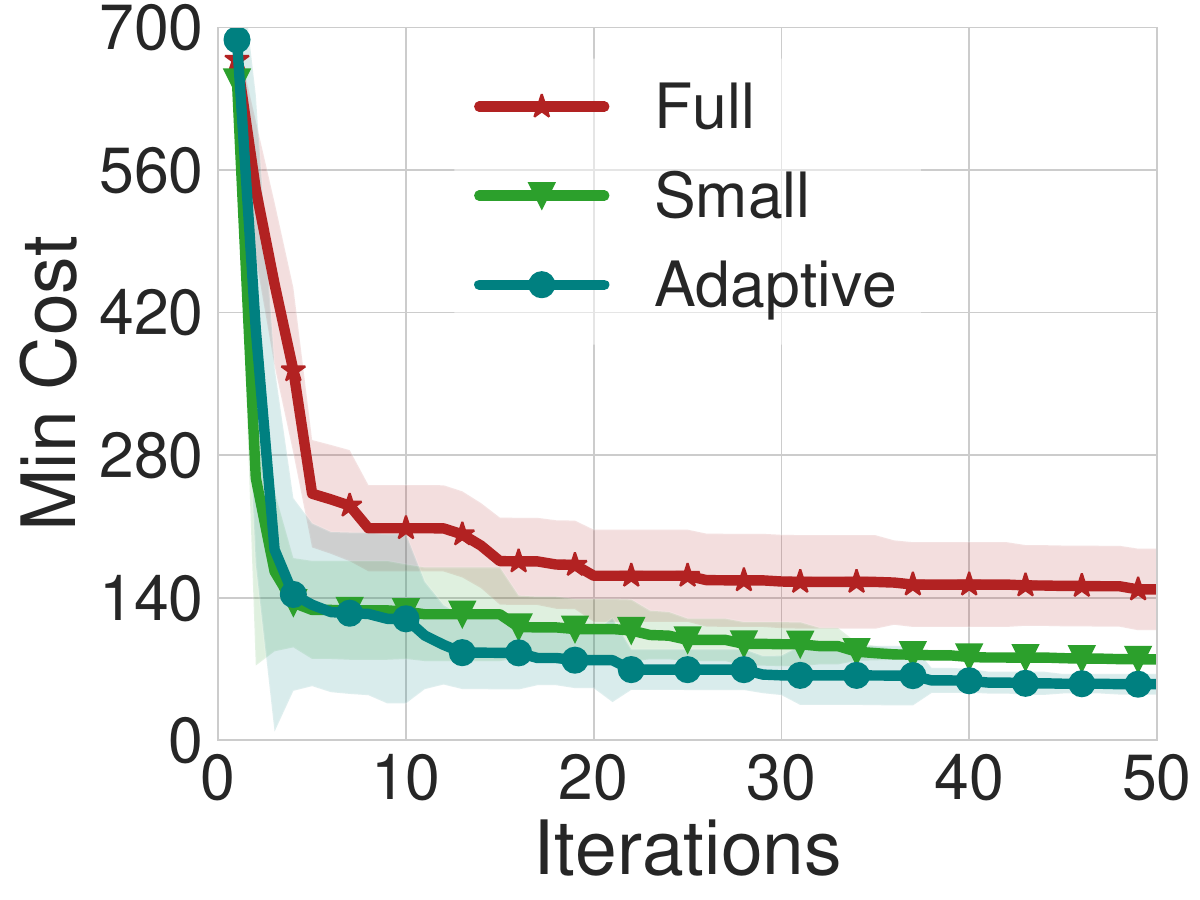}
            \label{fig:subspace:b}
	}}
	\caption{Left: Results when tuning PageRank and TeraSort over different sub-spaces compared with default configurations. Right: Optimization curve on Terasort.  }
  \label{fig:exp_space}
\end{figure}

\noindent
{\bf Acceleration with ensemble surrogate. }
Figure~\ref{fig:exp_metalearning_surrogate} shows the execution cost on Wordcount and Teresort with and without the ensemble surrogate, which carries meta-knowledge learned from related tasks to speed up the tuning process. 
We observe a clear reduction of the average cost over vanilla BO in the beginning 10 iterations. 
Concretely, BO with the ensemble surrogate takes at least three times fewer iterations to achieve the same average cost as vanilla BO using 30 iterations.

\subsection{Ablation Studies}
\label{sec:ablations}

In this section, we present additional analysis on three methods in Section~\ref{sec:config_acquisition}. Here the meta-learning module is disabled.

\noindent
{\bf Sub-space Generation. }
Figure~\ref{fig:exp_space} shows the evaluation costs when tuning PageRank and TeraSort with different configuration spaces.
We compare tuning with full space (30 parameters), small space (6 parameters with the highest importance in Table~\ref{tab:parameter_importance}), and the proposed adaptively generated sub-space.
Figure~\ref{fig:subspace:a} presents the reduction ratio after 30 iterations relative to default settings while Figures~\ref{fig:subspace:b} plot the average cost during the optimization.
By introducing the sub-space, the average cost is consistently lower than optimization on the entire search space.
Concretely, small space works well on PageRank; our adaptive method can detect this signal based on intermediate tuning results, and shrink the space size to achieve a similar result.
However, although tuning converges quickly over the small space on Terasort, its final performance degenerates, where the small space dismisses the near-optimal configs.
As shown in Figure~\ref{fig:subspace:b}, our adaptive method can converge to a better config by gradually increasing the size of the subspace.

\begin{figure}[tb]
	\centering
	\subfigure[WordCount]{
		\scalebox{0.47}[0.47]{
			\includegraphics[width=1\linewidth]{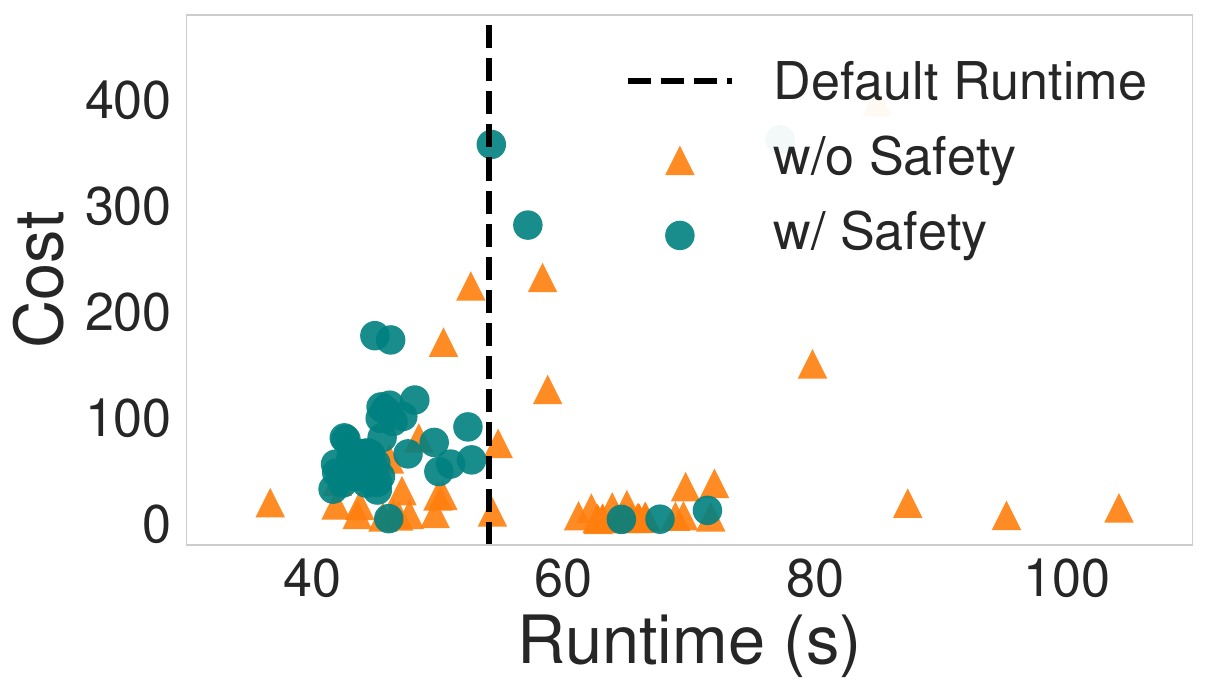}
	}}
	\subfigure[Bayes]{
		\scalebox{0.47}[0.47]{
			\includegraphics[width=1\linewidth]{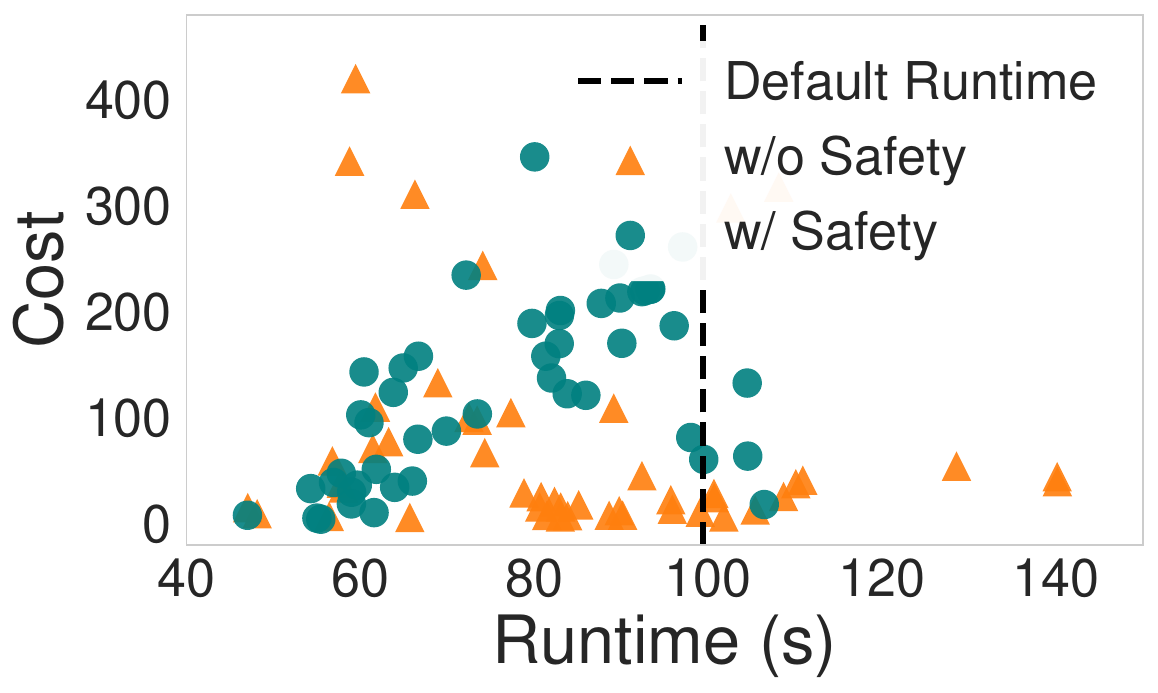}
	}}
	\caption{Analysis of safe exploration. Each point refers to the runtime (x-axis) and execution cost (y-axis) of each configuration during optimization. Configurations (circles or triangles) are infeasible on the right side of the dashed line.}
 \label{fig:exp_safety}
\end{figure}

\noindent
{\bf Safe Exploration and Exploitation. }
The empirical results on six Hibench tasks show that the average percentage of safe configurations generated by our method is $93.00\%$, which is much higher than vanilla BO's -- $69.67\%$.
Figure~\ref{fig:exp_safety} shows the objective (cost) and the constraint (runtime) during tuning WordCount and Bayes. 
We observe that, without the safety component, the ratio of infeasible configurations increases on both tasks.
Concretely, the safety component reduces the ratio of infeasible configurations from 56\% to 10\% and 20\% to 6\% on WordCount and Bayes, respectively.
In addition, the best objective found is slightly better on NWeight if we ignore the safety constraint.
The reason is that, our framework conservatively restricts the search space with the safe region so that the optimal objective value may be worse than that in the entire space.
However, while it is crucial to avoid infeasible configurations in online tuning tasks, it is essential to pursue safety at the expense of this insignificant objective degradation.

\begin{figure}[tb]
	\centering
		\scalebox{0.95}[0.95] {
		\includegraphics[width=1\linewidth]{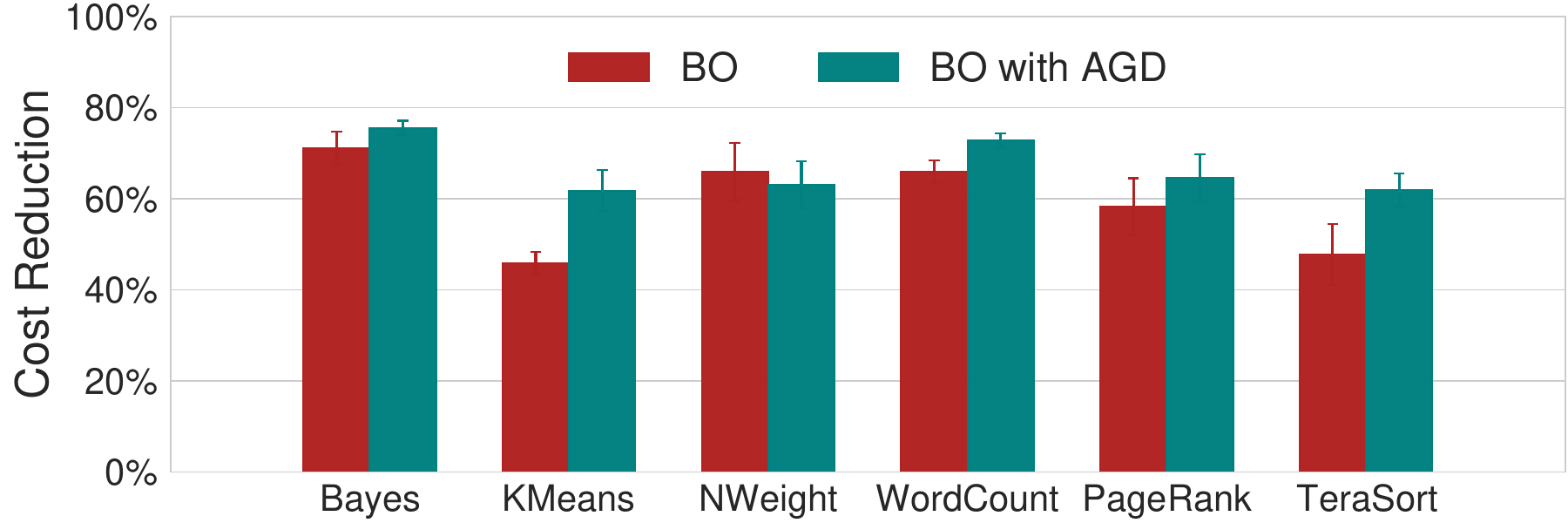}
         }
        \vspace{-0.5em}
	\caption{Tuning results when using approximate gradient descent relative to random search.}
    \label{fig:exp_grad}
\end{figure}

\noindent
{\bf Approximate Gradient Descent (AGD). }
Figure ~\ref{fig:exp_grad} shows the relative cost reduction with and without AGD on 6 HiBench tasks. 
Though BO with AGD shows a slight performance degradation on NWeight, it has a significant positive effect on other tasks.
Concretely, it further reduces the cost by 7.47\% on average relative to vanilla BO, which demonstrates its effectiveness.

\section{Related Work}

In the database community, many ML-based methods have been proposed to automatically tune the knobs in DBMS~\cite{DBLP:journals/pvldb/DuanTB09, fekry2020tuneful,DBLP:conf/sigmod/KunjirB20,DBLP:conf/sigmod/MaAHMPG18,zhang2022towards,zhang2021facilitating,zhang2023unified}, e.g., OtterTune~\cite{ottertune_demo,DBLP:conf/sigmod/AkenPGZ17}, Qtune~\cite{DBLP:journals/pvldb/LiZLG19}, CDBTune~\cite{DBLP:conf/sigmod/ZhangLZLXCXWCLR19}, ResTune~\cite{DBLP:conf/sigmod/ZhangWCJT0Z021} and HUNTER~\cite{cai2022hunter}.
When processing massive volumes of heterogeneous data, Spark tuning encounters special challenges compared with database tuning.
Considering a longer job runtime or a more expensive cost, Spark tuning requires as fewer trials as possible to finish the tuning process.
In addition, many researchers focus on resource management-related configuration optimization of cloud computing platforms~\cite{delimitrou2013paragon,delimitrou2014quasar,liu2016appbooster,zhu2021kea}. 
Here, we focus on tuning the Spark parameters efficiently.

The existing Spark tuning methods can be categorized into the following six classes~\cite{herodotou2020survey}:
(1) {\bf Rule-based methods}~\cite{spark_doc} require an in-depth understanding of complex system mechanisms and utilize the experience of human experts, online tutorials, or tuning instructions~\cite{spark_guide} to assist users to tune Spark parameters; the tuning process is often knowledge-intensive and labor-intensive.
(2) {\bf Cost modeling methods}~\cite{wang2015performance,venkataraman2016ernest,gounaris2017dynamic,singhal2017performance,zacheilas2017dione,chen2019cost} depend on a deep knowledge of system performance to build performance models.
However, it is difficult for them to capture complex system behaviors.
(3) {\bf Simulation-based approaches} use modular
or complete system simulation to conduct an execution with different parameter settings~\cite{de2018bigdatanetsim,de2019multi,karimian2019scalable,ardagna2021predicting}, and further build performance models to guide the tuning process.
These methods often fail to simulate complex internal dynamics in online tuning scenarios.
(4) {\bf Experiment-driven approaches}~\cite{bao2018learning,gounaris2017dynamic,petridis2016spark,yu2018datasize,zhu2017bestconfig} execute configuration evaluation, i.e., an experiment, repeatedly with different parameter configurations, and the configuration for each evaluation is suggested by a search algorithm.
Most of them are offline methods, which incur high overhead as they require additional evaluations in an offline cluster. 
(5) {\bf Machine learning (ML) approaches}~\cite{jia2016auto,wang2016novel,hernandez2018using,nguyen2018towards,chen2019d,prats2020you,zhu2021kea} employ ML techniques to build performance models, which need a large number of training samples with high overhead; owing to the advanced design in terms of convergence efficiency, our method does not require such many expensive training samples within the online tuning paradigm.
(6) {\bf Adaptive approaches}~\cite{ding2015jellyfish,bei2015rfhoc,genkin2016automatic,li2019qtune,fekry2020tuneful,kunjir2020black} tune parameters adaptively while an application is running, i.e., they can adjust the parameter settings online.
They need to modify the source code of Spark to achieve adaptive updates, while it incorporates additional costs and risks.
Besides, most of them focus on parameters in resource allocation (related to YARN containers) dynamically, while we concentrate on tuning all the Spark parameters.

\section{Conclusion}
In this paper, we presented a general and efficient online tuning framework for Spark, which can tune various objectives while satisfying application constraints. 
To solve the generalized tuning problem while preserving safety, we designed a Bayeisian optimization (BO) based solution along with a safe configuration acquisition approach.
To further accelerate the configuration search, we developed three innovative techniques within BO: adaptive sub-space generation, approximate gradient descent, and meta-learning-based knowledge transfer.
We implemented this framework as a tuning service and applied it to Tencent's data platform.
The empirical results on extensive benchmark and production tasks demonstrate its superior performance. 
In the future, we plan to extend this framework to support more data analytics systems.

\begin{acks}
We thank the engineers from the Data Platform department, Technology and Engineering Group (TEG) of Tencent for their technical and engineering support.
\end{acks}

\balance

\bibliographystyle{ACM-Reference-Format}
\bibliography{reference}


\end{document}